\documentclass[twocolumn,prb,reprint,showpacs,preprintnumbers,amsmath,amssymb,groupedaddress]{revtex4-1}
\usepackage{graphicx}
\usepackage{dcolumn}
\usepackage{bm}
\usepackage{graphicx}
\usepackage{subfigure}
\usepackage{physics}
\usepackage{epstopdf}
\usepackage[utf8]{inputenc}
\usepackage[english]{babel}
\usepackage{enumerate}
\usepackage{mathtools}
\usepackage{graphicx}
\usepackage{dcolumn}
\usepackage{bm}
\usepackage{graphicx}
\usepackage{subfigure}
\usepackage{physics}
\usepackage{epstopdf}
\usepackage[utf8]{inputenc}
\usepackage[english]{babel}
\usepackage{amsmath}
\usepackage{amssymb}
\usepackage{graphicx}
\usepackage{relsize}
\usepackage{xfrac}
\usepackage{epstopdf}
\usepackage{enumerate}
\usepackage{mathtools}

\usepackage[usenames, dvipsnames]{color}

\begin{document}

\title{Manipulation of non-linear heat currents in the dissipative Anderson-Holstein model}

\author{Bitan De}
\author{Bhaskaran Muralidharan}%
 \email{bm@ee.iitb.ac.in}
\affiliation{Department of Electrical Engineering, Indian Institute of Technology Bombay, Powai, Mumbai-400076, India
}%




\date{\today}

\begin{abstract}
\indent The anomalous behavior of electron induced phonon transport is investigated using an \textit{Anderson-Holstein} based dissipative quantum dot setup under two relevant bias situations: (a)  a voltage bias in the absence of an electronic temperature gradient and (b) an electronic temperature gradient at zero voltage.  It is shown that the direction of phonon transport in the non-linear regime is different in the two cases since the first case facilitates the accumulation of phonons in the dot and the second case leads to the absorption of phonons in the dot. In the linear regime, both the phonon and electronic transport get decoupled and Onsager's symmetry is verified.  We explain the observed cumulative effects of voltage and electronic temperature gradients on the non-linear phonon currents by introducing a new transport coefficient that we term as the \textit{electron induced phonon thermal conductivity}. It is demonstrated that under suitable operating conditions in Case (a) the dot can pump in phonons into the hotter phonon reservoirs and in Case (b) the dot can extract phonons out of the colder phonon reservoirs. Finally, we elaborate on how the non-linear electronic heat current can be stimulated and controlled by engineering the temperature of the phonon reservoirs even under vanishing effective electron flow.   

\end{abstract}
\maketitle


\section{\label{sec:level1}Introduction}
\indent The physics of electron-phonon interaction has gained significant interest \cite{elphrmp} with its fingerprints being captured in diverse physical systems like superconductors\cite{sc1,sc2,sc3,sc4,sc5}, cavity-coupled mesoscopic conductors\cite{cav1,cav2}, optomechanical systems\cite{opto1,opto2} and electronic devices\cite{device1,device2}, to name a few. So far, the research in this area is primarily carried out along two predominant directions: (a) exploration of novel experimental techniques to detect such interactions \cite{spec1,spec2,spec3,spec4,Ilani} and (b) theoretical studies on the effect of electron-phonon coupling on quantum phenomena in the nanoscale \cite{Hartle2011,Leijnse2010,Galperin2004,Galperin2007,de,bijoy1,oppen1,Sothmann2014,Siddiqui2006,Mitra2004}. Typically in microscopic thermoelectric setups, the former direction concerns materials synthesis, device fabrication and the integration at the systems level \cite{Kim2014,Reddy1568}. The principal goal in this case, is to control the flow of waste heat to optimize the efficiency of energy conversion\cite{science1,science2,science3,science4}. On the other hand, the latter direction deals with the dynamics of charge and heat transport from a quantum transport perspective. From the application point of view, this approach aims to propose new devices to realize novel paradigms like \textit{phonon computation} \cite{Jiang,Bli1,Bli2,Bli3,Karl} .\\  
\indent Many notable works have established that in the zero-dimensional systems such as quantum dots,\cite{Mahan1996,Kubala2006,Kubala2008} the interplay of electrons and phonons gives rise to some unique features manifesting as transport signatures \cite{Park2000,Leroy2004,LeRoy2005,Yu2004,Sapmaz2006}. Here, typical to quantum transport set ups, electronic and phonon currents are driven by applying a bias voltage and/or a thermal bias across the contacts. While voltage induced phonon transport has been the subject of many theoretical works\cite{ah1,ah2,ah3,ah4,ah5,ah6}, understanding the non-equilibrium phonon transport under an electronic temperature bias has not been explored so far, which will be the objective of this paper.

\begin{figure*}[]
	\begin{center}
		\subfigure[]{\includegraphics[width=0.45\textwidth, height=0.25\textwidth]{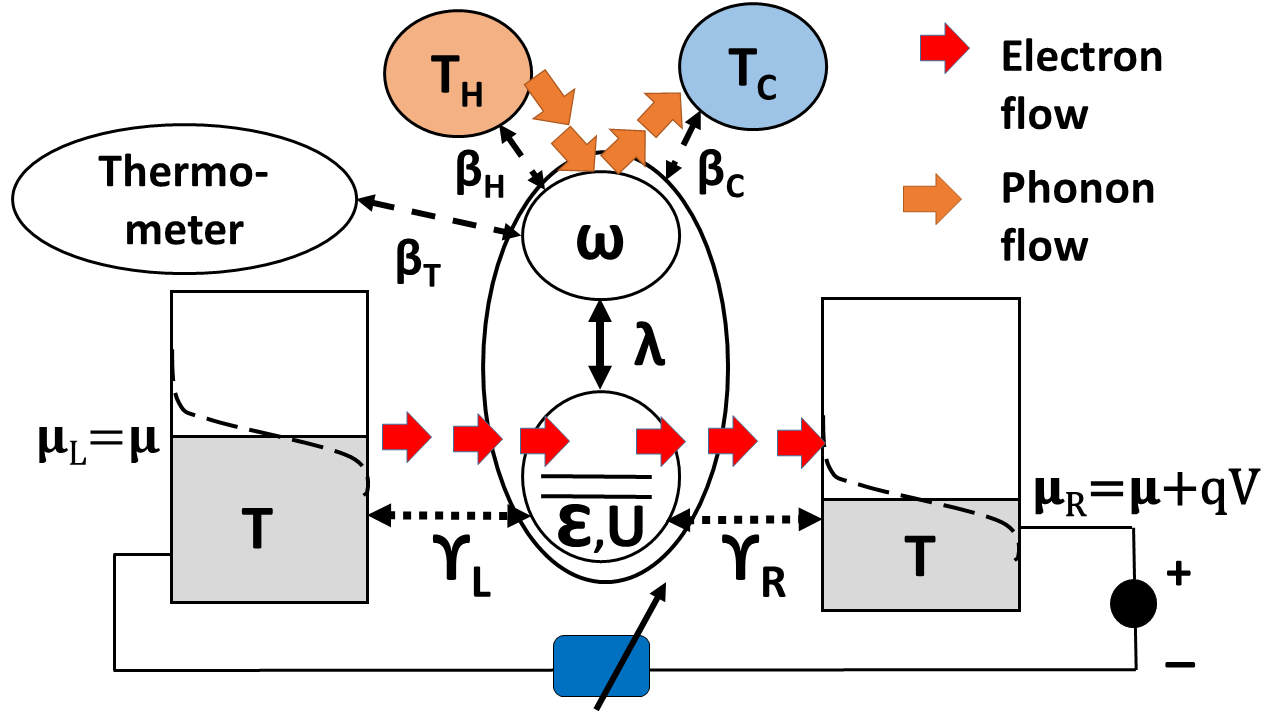}\label{1a}}
		\quad
		\subfigure[]{\includegraphics[width=0.45\textwidth, height=0.25\textwidth]{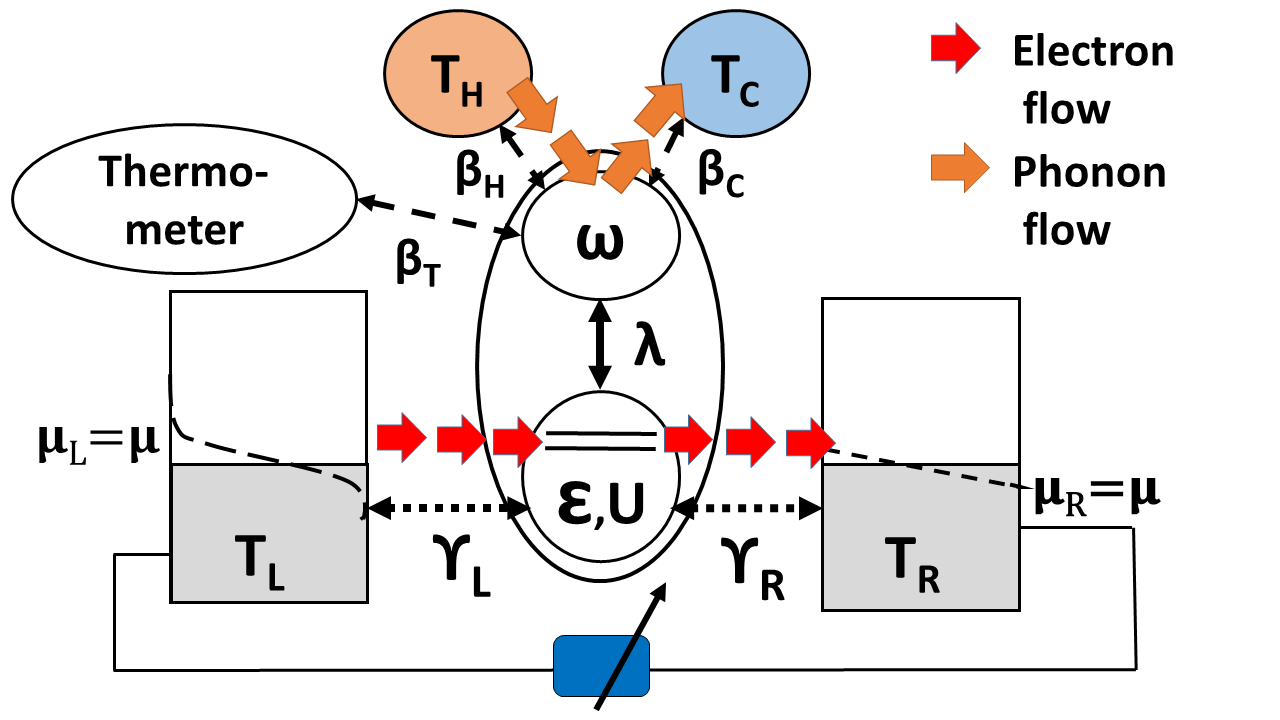}\label{1b}}
		\quad
		\subfigure[]{\includegraphics[width=0.45\textwidth, height=0.25\textwidth]{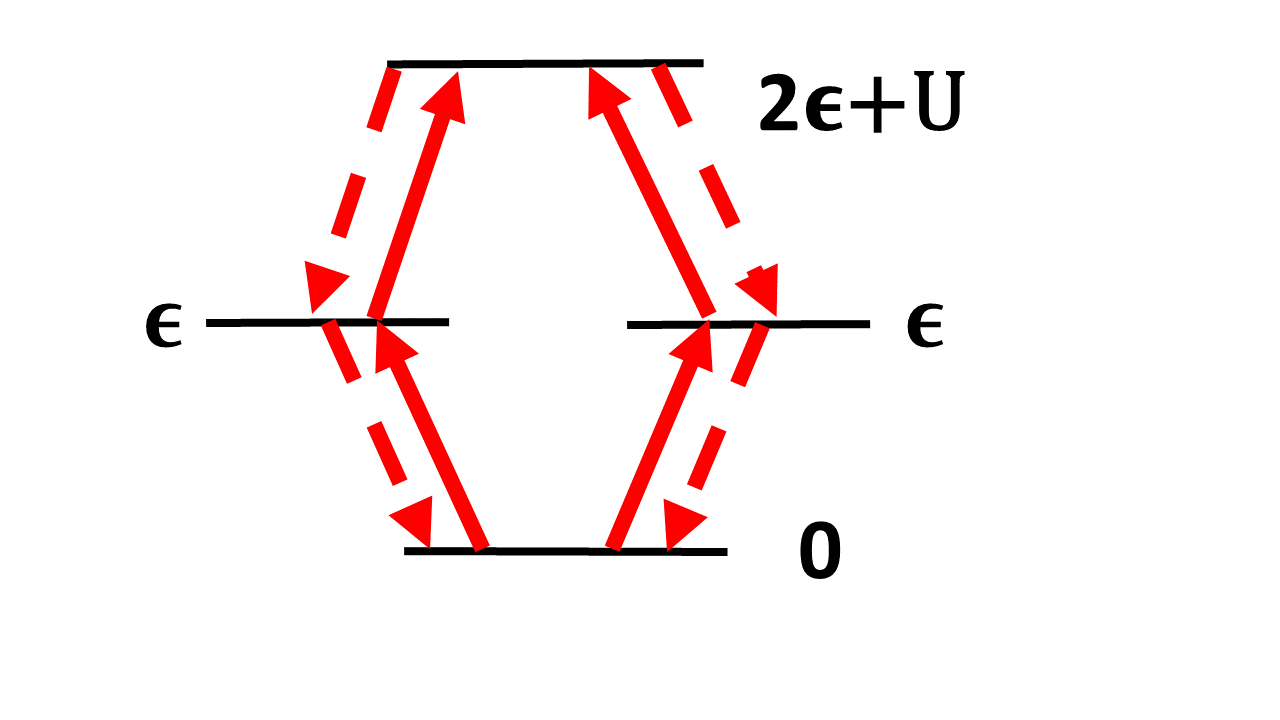}\label{1c}}
		\quad
		\subfigure[]{\includegraphics[width=0.45\textwidth, height=0.25\textwidth]{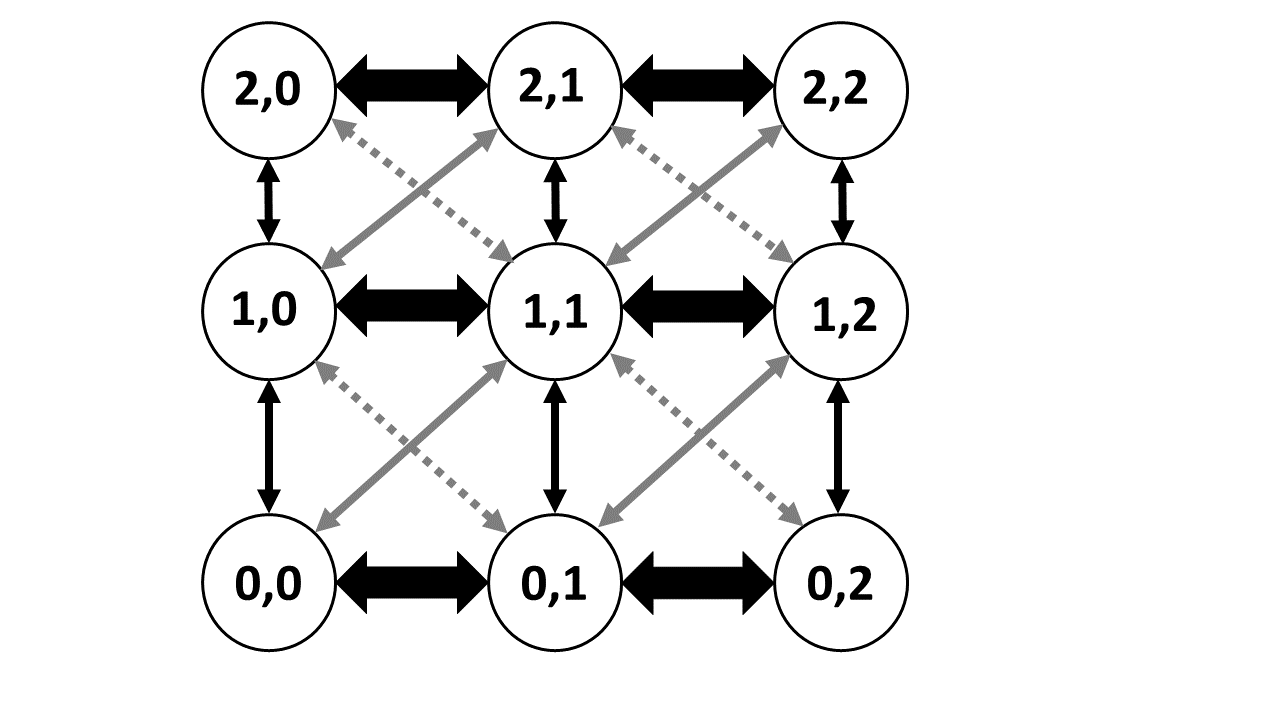}\label{1d}}
		
	\end{center}
	
	\caption{Non-equilibrium charge current(shown in red bold arrow) is driven through the central dot in two ways: either by applying (a) voltage bias or (b) by applying electronic temperature gradient across the contacts $L$ and $R$. Dot phonon current (shown by orange bold arrows) is set up by creating a thermal gradient across the reservoirs $H$ and $C$. In both (a) and (b) the dot temperature is monitored by \textit{phonon thermometer} bath. (c) The state transition diagram in the electronic Fock space. (d) The state transition diagram in the many body electron-phonon Fock space. Direct electron tunnelings (depicted by thin black double arrows) occur between states $|n,q\rangle$ and $|n\pm 1,q\rangle$, while phonon assisted electron tunnelings (depicted by gray thin double arrows) happen between states $|n,q\rangle$ and $|n\pm 1,q'\rangle$. Reservoir assisted phonon transitions (represented by black bold arrows) take place between the states $|n,q\rangle$ and $|n,q\pm 1\rangle$.}
	\label{fig1}
\end{figure*}
\indent  Our objective in this work is to compare the nature of the electronically stimulated phonon transport in two cases via: Case (a): a voltage bias in the absence of an electronic temperature bias and Case (b): an electronic temperature bias at zero voltage. It is shown by the simulation framework that the polarity of phonon current in the nonlinear regime is opposite in the two cases since the first case facilitates the accumulation of phonons in the dot and the second case leads to the absorption of phonons in the dot. \\
\indent In Case (a), surplus phonons heat up the dot and are suitably extracted by the phonon reservoirs kept in equilibrium. In Case (b), the dot cools down for being phonon deficient and the reservoirs pump in phonons into it. However in the linear regime, phonon and electron transport get decoupled and Onsager's reciprocity is verified. \\
\indent In our previous work \cite{nsr}, we explored the \textit{non-linear phonon Peltier effect} by noticing the voltage dependence of the non-linear phonon current. In the current work, we propose a new parameter called, \textit{electron-assisted phonon thermal conductivity} to signify the mutual dependence of phonon currents on the voltage and electronic temperature gradient. \\
\indent We also notice that the polarity of phonon transport can be controlled by differing the temperature of phonon reservoirs from the temperature of electronic contacts. In Case (a), phonons flow from the reservoirs to the dot at low voltage when the reservoirs are hotter than the contacts. At large voltage, the polarity of phonon flow is reversed and the phonons are pumped into the hotter reservoirs. An opposite situation is observed in Case (b) given the reservoirs are colder than the contacts. In this case, phonons relax from the dot to the reservoirs at low electronic temperature gradient. When the temperature gradient is large, the direction of phonon flow is reversed and phonons are extracted from the colder reservoirs. In both cases, the direction of phonon current is verified by estimating the dot temperature with a phonon thermometer bath weakly coupled to the dot. Finally, we elaborate on how the non-linear electronic heat current can be stimulated and controlled by engineering the temperature of the phonon reservoirs even even under vanishing effective electron flow.\\
\indent This paper is organized as follows: Section II introduces the device model and formulates the transport equations. In the Sec. III A,  we focus on the comparative study of the characteristics of the phonon currents in the two cases under consideration. Section III B  introduces the electron assisted phonon thermal conductivity coefficient and compares it with the conventional electronic thermal conductivity. In the Sec. III C,  we describe the engineering of the temperature of the phonon reservoirs to modulate the flow of the phonon and electronic heat currents. In the Sec. IV, we summarize our results and conclude.


\section{\label{sec:level2}Physics and Formulation}
\indent The schematics of the set ups to be studied here are depicted in Fig.\ref{1a} and Fig.\ref{1b}. Each of them typically comprises a single quantum dot weakly coupled to the electronic contacts $\alpha_1$ ($\alpha_1 \in L, R$), and the phonon reservoirs $\alpha_2$ ($\alpha_2 \in H, C$). The dot is described by the dissipative \textit{Anderson-Holestein} model which has both electronic and phonon degrees of freedom interacting with each other. The electronic current is driven through the device in two ways: Case (i): by applying a voltage bias (shown in Fig.\ref{fig1}(a)) across the contacts $L$ and $R$ and Case (ii) by applying an electronic temperature gradient (shown in Fig.\ref{1b}) across the contacts $L$ and $R$ . The phonon current is set up by applying a thermal bias across the phonon reservoirs $H$ and $C$. In addition, the electron current gives rise to a phonon current even in the absence of a temperature bias across the reservoirs $H$ and $C$, provided the electron-phonon coupling is finite. The direction of phonon flow depends on the dot temperature ($T_M$), which is estimated by a phonon thermometer bath weakly coupled to the dot.
\subsubsection{Model Hamiltonian}
\indent The composite system Hamiltonian $\hat{H}$ can be written as $\hat{H}=\hat{H}_D+\hat{H}_{\alpha_1}+\hat{H}_{\alpha_2}+\hat{H}_{\alpha_1 D}+\hat{H}_{\alpha_2 D}$. Here $\hat{H}_{D}$, $\hat{H}_{\alpha_1}$, $\hat{H}_{\alpha_2}$ denote the respective Hamiltonians of the dot, the contacts and the reservoirs, while $\hat{H}_{\alpha_1 D}$ and $\hat{H}_{\alpha_2 D}$ represent the dot-to-contact electronic tunneling processes and the dot-to-reservoir phonon relaxation processes respectively. The dot Hamiltonian is further divided as,
\begin{equation}
\begin{gathered}
\hat{H}_{D}=\hat{H}_{el}+\hat{H}_{ph}+\hat{H}_{el-ph}.
\end{gathered}
\label{Eq1}
\end{equation}
The electronic part  ($\hat{H}_{el}$) consists of a single spin degenerate energy level with an on-site energy $\epsilon$ and a Coulomb interaction energy $U$ for double occupancy. The phonon part ($\hat{H}_{ph}$) comprises of a single phonon mode $\nu$ with angular frequency $\omega_{\nu}$. Inside the dot, the electrons and phonons interact through a dimensionless parameter $\lambda_{\nu}$. Hence, $H_{el}$, $H_{ph}$,$H_{el-ph}$ are given by, 
\begin{equation}
\begin{gathered}
\hat{H}_{el}=\bigg(\sum\limits_{\sigma}^{}\epsilon d_{\sigma}^{\dagger}d_{\sigma}+U d_{\uparrow}^{\dagger}d_{\uparrow}d_{\downarrow}^{\dagger}d_{\downarrow}\bigg),\\
\hat{H}_{ph}=\hbar \omega_{\nu}\hat{b}_{\nu}^{\dagger}\hat{b}_{\nu},\\
\hat{H}_{el-ph}=\sum\limits_{\sigma}^{}\lambda_{\nu}\hbar\omega_{\nu}d_{\sigma}^{\dagger}d_{\sigma}(\hat{b}_{\nu}^{\dagger}+\hat{b}_{\nu}).
\end{gathered}
\label{Eq2}
\end{equation}
In the above expressions, $d_{\sigma}^{\dagger}$($d_{\sigma}$) and $b_{\nu}^{\dagger}$($b_{\nu}$) are the respective creation(annihilation) operators of the electrons and phonons. The electronic contacts $\alpha_1$ characterize a macroscopic body of non-interacting electrons in the momentum eigen state $\alpha_1 k$ and spin $\sigma'$ with energies $\epsilon_{\alpha_1 k\sigma'}$. Similarly, the phonon reservoirs $\alpha_2$ comprise of numerous phonon modes $\alpha_2 r$ with an angular frequency $\omega_{\alpha_2 r}$. They  are defined by the following Hamiltonians,
\begin{equation}
\begin{gathered}
\hat{H}_{\alpha_1}=\sum\limits_{\alpha_1\in L,R}^{}\sum\limits_{\alpha_1 k\sigma'}^{}\hat{c}_{\alpha_1 k\sigma'}^{\dagger}\hat{c}_{\alpha_1 k\sigma'},\\
\hat{H}_{\alpha_2}=\sum\limits_{\alpha_2\in H,C}^{}\sum\limits_{\alpha_2}^{}\hat{B}_{\alpha_2r}^{\dagger}\hat{B}_{\alpha_2r},
\end{gathered}
\label{Eq3}
\end{equation}
where $\hat{c}_{\alpha_1 k\sigma'}^{\dagger}(\hat{c}_{\alpha_1 k\sigma'})$ creates (annihilates) an electron with momentum $k$ and spin $\sigma'$ in the contact $\alpha_1$, and $\hat{B}_{\alpha_2 r }^{\dagger}(\hat{B}_{\alpha_2 r })$ creates (annihilates) a phonon of angular frequency $\omega_{\alpha_2 r}$ in the reservoir $\alpha_2$. If the electrons in the dot are coupled to the electrons in the contacts $\alpha_1$ through an energy $\tau_{\alpha_1 k\sigma' \sigma}^{el}$ and the phonons in the dot are coupled to the phonons in the reservoirs $\alpha_2$ through an energy $\tau_{\alpha_2r\nu}^{ph}$ then the coupling Hamiltonians $\hat{H}_{\alpha_1 D}$ and $\hat{H}_{\alpha_2 D}$ are given as follows
\begin{equation}
\begin{gathered}
\hat{H}_{\alpha_1 D}=\sum\limits_{k\sigma'\sigma}^{}[\tau_{\alpha_1 k\sigma' \sigma}^{el}\hat{c}_{\alpha_1 k\sigma'}^{\dagger}\hat{d}_{\sigma}+H.c],\\
\hat{H}_{\alpha_2 D}=\sum\limits_{\nu,\alpha_2}^{}\tau_{\alpha_2r\nu}^{ph}(\hat{B}_{\alpha2r}^{\dagger}+\hat{B}_{\alpha2r})(\hat{b}_{\nu}^{\dagger}+\hat{b}_{\nu}).
\end{gathered}
\label{Eq4}
\end{equation} 
In this work, we assume that the phonon transition processes are mode independent and the electron tunneling processes are momentum conserved and spin independent. With this approximation we can  rewrite $\tau_{\alpha_1 k\sigma' \sigma}^{el}$ and $\tau_{\alpha_2r\nu}^{ph}$ simply as $\tau^{el}$ and $\tau^{ph}$ respectively.\\
\indent The dot Hamiltonian is diagonalized by the polaron transformation (such that $\hat{H}_{D}\rightarrow\tilde{\hat{H}}_D=e^S\hat{H}_De^{-S}$, where $S=\sum_{\nu}^{}\lambda_{\nu}[\hat{b}_{\nu}-\hat{b}_{\nu}^{\dagger}$]). It transforms the dot fermionic operators $\hat{d}_{\sigma}^{\dagger}$($\hat{d}_{\sigma}$) and leads to the renormalization of $\epsilon$ and $U$, such that $\tilde{\epsilon}=\epsilon-\lambda_{\nu}^2\hbar\omega_{\nu}$ and  $\tilde{U}=U-2\lambda_{\nu}^2\hbar\omega_{\nu}$. The eigen states of $\tilde{\hat{H}}_{D}$ are represented by $|n,q\rangle$, where $n$ and $q$ are the electron and phonon number of the eigen states. The energy eigen values of those states become $E_{(n,q)}=\tilde{E}_n+q\hbar\omega_{\nu}$, where, $\tilde{E}_{0}=0,\tilde{E}_{1}=\tilde{\epsilon},\tilde{E}_{2}=2\tilde{\epsilon}+\tilde{U}$, corresponding to the $n=0,1$ and $2$ electron number spaces respectively. This transformation also renormalizes the electronic tunneling energy $\tau^{el}$, such that $\tilde{\tau}^{el}=\tau^{el}\exp[-\lambda_{\nu}(\hat{b_{\nu}}-\hat{b_{\nu}}^{\dagger})]$. However, the phonon coupling energy $\tau^{ph}$ is left unaltered due to the polaron transformation since the operator $S$ commutes with the phonon operators $b_{\nu}^{\dagger}$($b_{\nu}$) of the dot .\\
\indent  With the derived expressions of $\tilde{\tau}_{el}$ and $\tau_{ph}$, we can evaluate the dot-to-contact electron tunneling rate $\gamma_{\alpha_1}$  and the dot-to-phonon relaxation rate $\beta_{\alpha_2}$ using the Fermi's golden rule. They are represented as:  $\gamma_{\alpha_1}=\frac{2\pi}{\hbar}\sum_{\alpha_1}^{}\abs{\tilde{\tau}_{\alpha_1}^{el}}^2\rho_{\alpha_1\sigma'}$ and
$\beta_{\alpha_2}=\frac{2\pi}{\hbar}\sum_{\alpha_2}^{}\abs{\tau_{\alpha_2}^{ph}}^2D_{\alpha_2}$, where $\rho_{\alpha_1\sigma'}$ and $D_{\alpha_2}$ are the constant electron and phonon density of states associated with the contacts $\alpha_1$ and the reservoirs $\alpha_2$ respectively. The study of the model Hamiltonian enables us to formulate the transport methodology, which we will analyze in the next subsection.

\subsubsection{Transport methodology}
\indent Before we start the formulation of transport equations, it is imperative to explain the important approximations we have considered in this work. First, we set the rate of dot-to-reservoir phonon relaxation processes much lower than the rate of dot-to-contact electron tunneling processes ($\hbar\gamma_{\alpha_1}>>\hbar\beta_{\alpha_2}$) to rule out the system damping\cite{Braig2003}. In this limit, the phonon currents emanating from the terminal phonon reservoirs remain uncorrelated and hence can be computed independently\cite{Segal2005,Segal2006,Segall2005}. Second, we set $\hbar\gamma_{\alpha_1},\hbar\beta_{\alpha_2}<<k_{B}T$, so that the transport of the electrons and phonons through a single quantum dot weakly coupled to the electronic contacts and the phonon reservoirs can be described in the sequential tunneling regime in which charge and phonon currents are calculated via the quantum Master equation\cite{Basky_Beenakker,Basky_Datta,Beenakker,Basky_Milena,Muralidharan2012}. Last, we neglect the overlap of the adjacent phonon sidebands, by setting the energy gap between the sidebands larger than the tunneling induced broadening of energy levels ($\hbar\omega_{\nu}>>\hbar\gamma_{\alpha_1}$)\cite{Timm}. With this assumption, the Markov approximation is justified and two consecutive electron tunneling processes are completely uncorrelated since the  memory of macroscopic contacts and reservoirs is negligible\cite{Theses1}. Hence, the diagonal terms of the system density matrix get decoupled from the off-diagonal terms and the Master equation reduces to the rate equation\cite{Brouw,Koenig_1,Koenig_2}.  \\
\indent The tunneling rate between two eigen states $|n,q\rangle$ and $|n\pm 1,q'\rangle$ is determined by the contact Fermi function of the energy difference of the two states and it is given by:
\begin{equation}
\begin{gathered}
R_{(n,q)\rightarrow(n+1,q')}=\sum\limits_{\alpha_1\in L,R}^{}\gamma_{\alpha_1}|\langle n,q|\hat{d}_{\sigma}|n+1,q'\rangle|^2 \\ \times f_{\alpha_1}(E_{n+1,q'}-E_{n,q})
\end{gathered},
\label{Eq5}
\end{equation}
\begin{equation}
\begin{gathered}
R_{(n,q)\rightarrow(n-1,q')}=\sum\limits_{\alpha_1\in L,R}^{}\gamma_{\alpha_1}|\langle n,q|\hat{d}_{\sigma}^{\dagger}|n-1,q'\rangle|^2\\ \times [1-f_{\alpha_1}(E_{n,q}-E_{n-1,q'})],
\end{gathered}
\label{Eq6}
\end{equation}
where $f_{\alpha_1}(\zeta)=1/(1+exp(\frac{\zeta-\mu_{\alpha_1}}{k_B T_{\alpha_1}}))$ is the Ferni-Dirac distribution function of the contact $\alpha_1$ with chemical potential $\mu_{\alpha_1}$ and temperature $T_{\alpha_1}$. The phonon relaxation process between the reservoirs and the dot lead to the transition between two eigen states $|n,q\rangle$ and $|n,q\pm 1\rangle$ and the relaxation rate follows the Boltzmann ratio:
\begin{equation}
\begin{split}
R_{(n,q)\rightarrow(n,q+1)}=\sum\limits_{\alpha_2\in H,C}^{}\beta_{\alpha_2}(q+1)exp\bigg(-\frac{\hbar\omega_{\nu}}{k_{B}T_{\alpha_2}}\bigg),
\end{split}
\label{Eq7}
\end{equation}
\begin{equation}
\begin{split}
R_{(n,q)\rightarrow(n,q-1)}=\sum\limits_{\alpha_2\in H,C}^{}\beta_{\alpha_2}(q+1).
\end{split}
\label{Eq8}
\end{equation}
Using the rate equations, the master equation for the probabilities $P_{(n,q)}$ of the many-body electron-phonon states $|n,q\rangle$ take the following form:
\begin{equation}
\begin{gathered}
\frac{dP_{(n,q)}}{dt}=\sum\limits_{q'}^{N_q}\bigg[R_{(n',q')\rightarrow(n,q)}^{el}P_{(n',q')}-R_{(n,q)\rightarrow(n',q')}^{el}P_{(n,q)}\bigg]\\+\bigg[R_{(n',q')\rightarrow(n,q)}^{ph}P_{(n',q')}-\\R_{(n,q)\rightarrow(n',q')}^{ph}P_{(n,q)}\bigg]\delta(n\pm 1,n')\delta(q\pm 1,q').
\end{gathered}
\label{Eq9}
\end{equation}
In the steady state, the derivative in the left hand side of \eqref{Eq9} vanishes and the steady state probabilities  $P_{(n,q)}$ can be determined by solving the algebric equations. We use the probabilities to compute the charge currents, the electronic heat currents associated with the contacts $\alpha_1$ and the phonon heat currents associated with reservoirs $\alpha_2$. They are given as,
\begin{equation}
\begin{gathered}
I_{\alpha_1}=\sum\limits_{q=0}^{N_q}\sum\limits_{q'=0}^{N_q} -q\bigg[R_{(n+ 1,q')\rightarrow(n,q)}^{el_{\alpha_1}}P_{(n+1,q')}\\-R_{(n+ 1,q')\rightarrow(n,q)}^{el_{\alpha_1}}P_{(n,q)}\bigg],
\end{gathered}
\label{Eq10}
\end{equation}
\begin{equation}
\begin{gathered}
I_{el_{\alpha_1}}^Q=\sum\limits_{q=0}^{N_q}\sum\limits_{q'=0}^{N_q} (E_{n+ 1,q'}-E_{n,q}-\mu_{\alpha})\\ \bigg[R_{(n+1,q')\rightarrow(n,q)}^{el_{\alpha_1}}P_{(n+ 1,q')}\\ -R_{(n+1,q')\rightarrow(n,q)}^{el_{\alpha_1}}P_{(n,q)}\bigg],
\end{gathered}
\label{Eq11}
\end{equation}
\begin{equation}
\begin{gathered}
I_{ph_{\alpha_2}}^Q=\sum\limits_{q=0}^{N_q}\sum\limits_{q'=0}^{N_q} \hbar\omega_{\nu} \bigg[R_{(n,q)\rightarrow(n',q')}^{ph_{\alpha_2}}P_{(n,q)}\\ -R_{(n',q')\rightarrow(n,q)}^{ph_{\alpha_2}}P_{(n',q')}\bigg]\delta(n,n')\delta(q\pm 1,q').
\end{gathered}
\label{Eq12}
\end{equation}
Now it is evident from \eqref{Eq5} and \eqref{Eq6} that the transition between the states $|n,q\rangle$ and $|n\pm 1,q'\rangle$ leads to the net phonon generation (or absorption) in the dot. In the next subsection we will focus on the role of $\lambda_{\nu}$ in controlling the phonon generation(or absorption). Considering the law of charge conservation, in the rest of the paper we will denote $I_{L}=-I_{R}=I$.
\subsubsection{Effect of interaction on phonon transport}
\indent The effective electron tunneling rate ($\gamma_{\alpha_1}^{eff}$) between the states $|n,q\rangle$ and $|n\pm 1,q'\rangle$ is modified by the \textit{Frank-Condon} overlapping factor between the two states\cite{Koch2005,Koch2006,Theses}. The effective electron tunneling rate is given as
\begin{equation}
\begin{gathered}
\gamma_{\alpha_1}^{eff}=\gamma_{\alpha_1}|FC_{q,q'}|^2=\gamma_{\alpha_1}\bigg[|\langle n,q|\hat{d}^{\dagger}|n',q'\rangle|^2\delta(n',n-1)\\+|\langle n,q|\hat{d}|n',q'\rangle|^2\delta(n',n+1)\bigg],
\end{gathered}
\label{Eq13}
\end{equation} 
where $|FC_{q,q'}|^2=exp(-\lambda_{\nu}^2)\frac{k!}{K!}\lambda^{2(K-k)}[L_k^{K-k}(\lambda_\nu^2)]^2$ is the \textit{Frank-Condon} factor between the two states with phonon number $q$ and $q'$ and $L_{k}^{K-k}$ is the associated Laguerre polynomial with $k=min(q,q')$ and $K=max(q,q')$. The net phonon generation (or absorption) in the dot due to the transition between the states $|n,q\rangle$ and $|n\pm 1,q\rangle$, takes place at a rate\cite{Siddiqui2006}
\begin{equation}
\begin{gathered}
GE_{ph}^{\alpha_1}=\sum\limits_{n,q}^{}\sum\limits_{n\pm 1}^{}(q'-q)P_{n,q}R_{(n,q)\rightarrow(n\pm 1,q')}^{\alpha_1},
\end{gathered}
\label{Eq14}
\end{equation}
 It is evident that in  the limit $\hbar\gamma_{\alpha_1}>>\hbar\beta_{\alpha_2}$, the phonon distribution in the dot is primarily determined by the electron transport. When $q\neq q'$, the phonons generate (or get absorbed) in the dot and the average phonon number in the dot $\langle N_{ph}\rangle=\sum\limits_{n,q}^{}qP_{n,q}$ deviates from the equilibrium phonon distribution $\langle N_{ph}^{eq}\rangle$ of the phonon reservoirs. The excess phonons are extracted  by the reservoirs at a rate\cite{Siddiqui2006} 
\begin{equation}
\begin{gathered}
\begin{split}
RE_{ph}^{\alpha_2}=\sum\limits_{q=0}^{N_q}\sum\limits_{q'=0}^{N_q} \hbar\omega_{\nu} \bigg[R_{(n,q)\rightarrow(n',q')}^{ph_{\alpha_2}}P_{(n,q)}\\ -R_{(n',q')\rightarrow(n,q)}^{ph_{\alpha_2}}P_{(n',q')}\bigg]\delta(n,n')\delta(q\pm 1,q')\\RE_{ph}^{\alpha_2}=\beta_{\alpha_2}\frac{\langle N_{ph}\rangle-\langle N_{ph}\rangle^{eq}}{1+\langle N_{ph}\rangle^{eq}}.
\end{split}
\end{gathered}
\label{Eq15}
\end{equation}
It is evident from \eqref{Eq12} and \eqref{Eq15} that the phonon current $I_{ph}^Q=\hbar \omega RE_{ph}^{\alpha_2}$. It is noticed from \eqref{Eq13}, that when $\lambda_{\nu}=0$, $\gamma_{\alpha_1}^{eff}$ vanishes unless $q=q'$. In this case, $\langle N_{ph} \rangle$ equals with $\langle N_{ph}^{eq} \rangle$ and $I_{ph_{\alpha_2}}^{Q}$ vanishes. On the other way, when $\lambda_{\nu}$ is non-zero, $\langle N_{ph} \rangle$ deviates from $\langle N_{ph}^{eq} \rangle$ and $I_{ph_{\alpha_2}}^{Q}$ becomes finite. The deviation of $\langle N_{ph}\rangle$ from $\langle N_{ph}^{eq}\rangle$ causes the variation of  dot temperature $T_M$  from the equilibrium reservoir temperature $T_{\alpha_2}$. The dot temperature can be computed from the Boltzmann ratio by with a quasi-equilibrium approximation\cite{oppen1} as:
\begin{equation}
\begin{gathered}
T_M=\frac{\hbar\omega_{\nu}}{k_B}\bigg[ln\bigg(\frac{P_{n,q}}{P_{n,q+1}}\bigg)\bigg]^{-1}
\end{gathered}
\label{Eq16}
\end{equation}
 We will notice in the next section that the electron induced phonon current shows different characteristics depending on whether the electron flow is stimulated by a voltage bias or by an electronic temperature bias. From now on, we will denote $\lambda_{\nu}$, $\omega_{\nu}$, $GE_{ph}^{\alpha_1}$ and $RE_{ph}^{\alpha_2}$ simply as $\lambda$, $\omega$, $G_{ph}$ and $R_{ph}$. Also, unless otherwise mentioned, the electronic contact coupling and reservoir phonon couplings are assumed to be symmetric (i.e $\gamma_{\alpha_1}=\gamma$, $\gamma_{\alpha_1}^{eff}=\gamma^{eff}$  and $\beta_{\alpha_2}=\beta$, where $\alpha_2 \in H,C$).\\
\begin{center}
	\begin{figure}[!htb]
		\subfigure[]{\includegraphics[width=0.225\textwidth, height=0.18\textwidth]{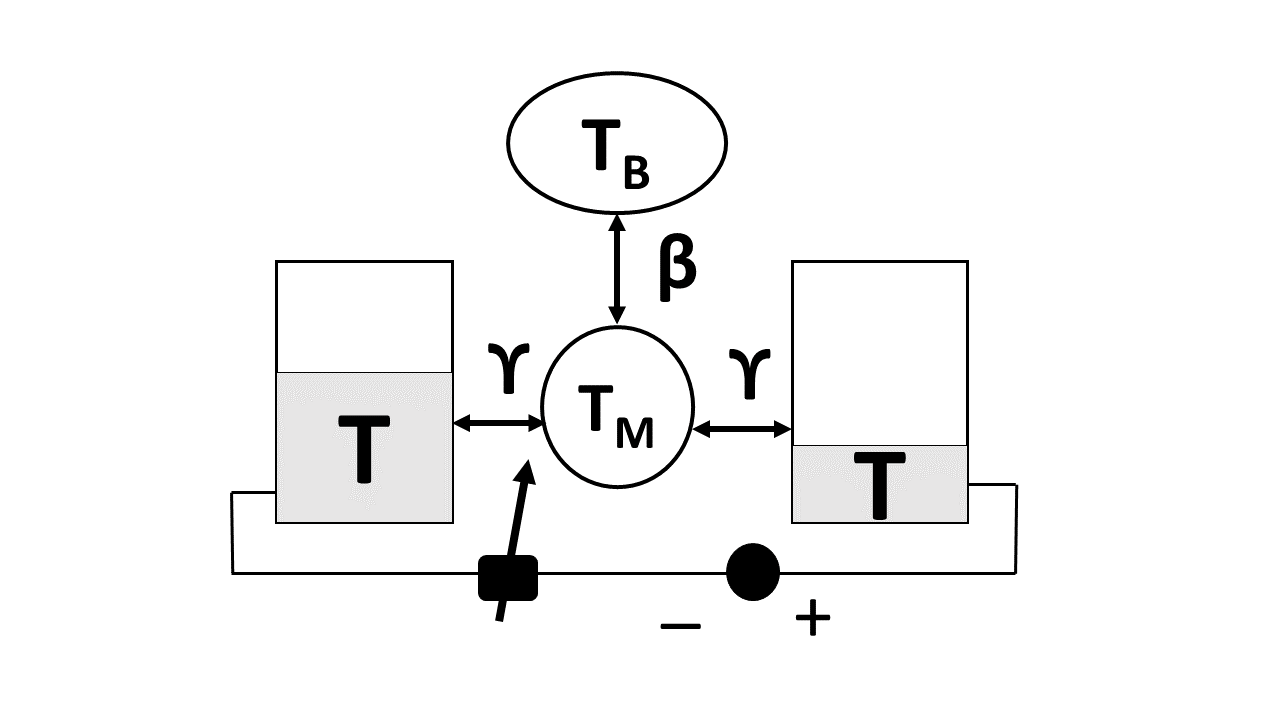}\label{2a}}
		\quad
		\subfigure[]{\includegraphics[width=0.225\textwidth, height=0.18\textwidth]{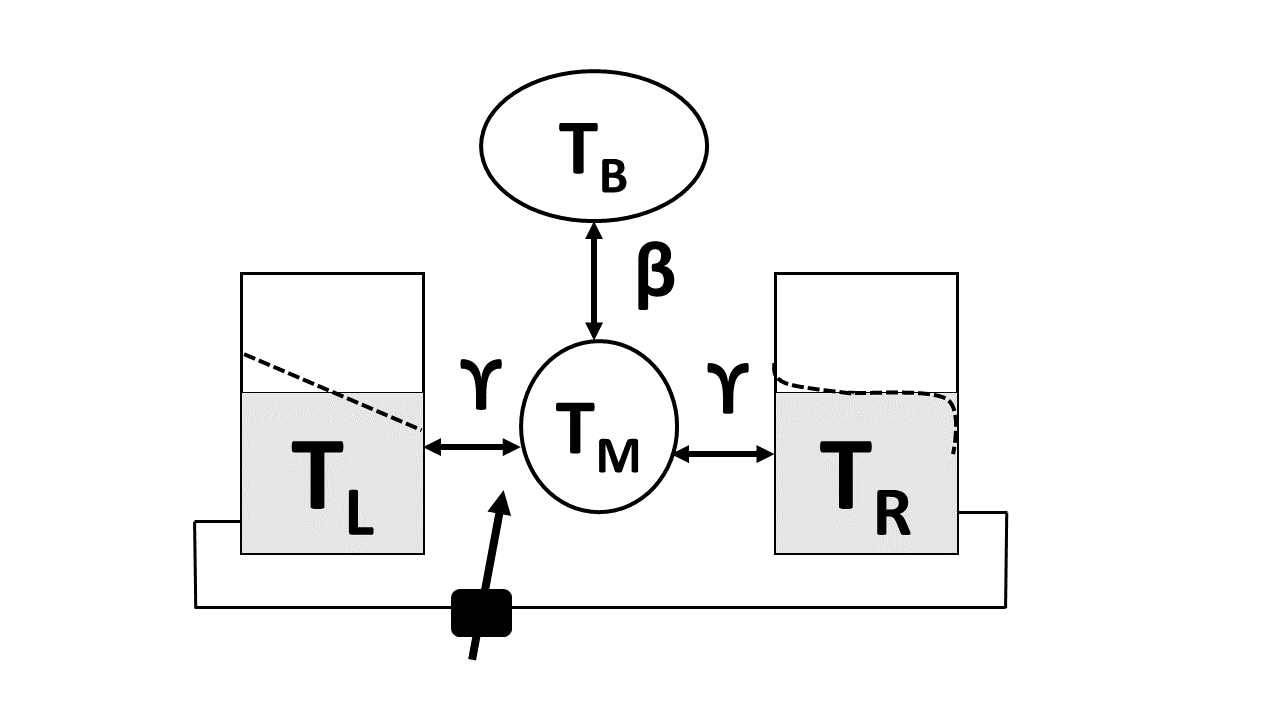}\label{2b}}
		\quad
		\subfigure[]{\includegraphics[width=0.225\textwidth, height=0.18\textwidth]{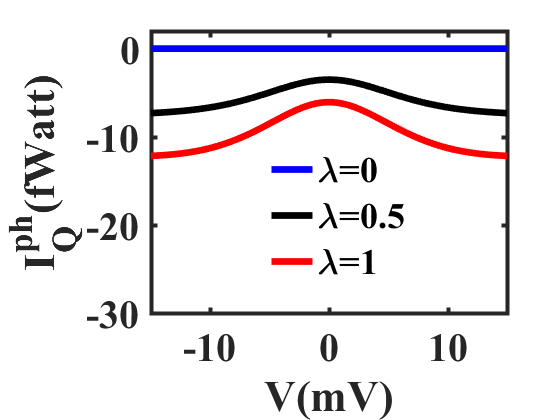}\label{2c}}
		\quad
		\subfigure[]{\includegraphics[width=0.225\textwidth, height=0.18\textwidth]{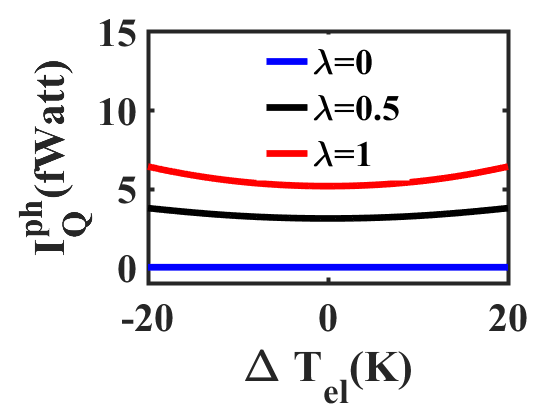}\label{2d}}
		\caption{Electron induced phonon transport at zero phonon temperature gradient. Modified setups in two operating limits: (a) Current is driven by a voltage bias when $\Delta T_{el}=0$. (b) Current is driven set up by an electronic temperature bias when $V=0$. (c) Variation of $I_Q^{ph}$ with voltage for different $\lambda$ in Case (a). (d) Variation of $I_Q^{ph}$ with $\Delta T_{el}$ for different $\lambda$ in Case (b). This figure establishes that $V$ and $\Delta T_{el}$ decide the direction of $I_Q^{ph}$ and $\lambda$ works like a phonon switch.}
		\label{fig2}
	\end{figure} 
\end{center}
\section{Results}
\subsection{Electron-phonon coupled transport }
\indent In this section we focus on the electron-phonon coupled transport in our setup. Each operating point is signified by a voltage bias, an electronic temperature gradient ($\Delta T_{el}=T_{L}-T_{R}$) applied across the contacts and a phonon temperature gradient ($\Delta T_{ph}=T_{H}-T_{C}$) applied between the reservoirs. In our analysis, we set ($k_{B}$T$>>\hbar \omega>>\hbar \gamma,\hbar \beta$) the tunnel induced broadening of the energy levels in the dot small to ensure that the levels do not overlap. Under this condition, transport is described in the sequential tunneling limit, where charge and heat current are calculated via rate equation. Additionally we assume the dot functions as n-type, i.e $\epsilon >>\mu$.\\
\indent First we demonstrate the effect of electron current on $I_Q^{ph}$ in two regimes: Case (a) the electron current is driven by a voltage bias when $\Delta T_{el}$ ($\Delta T_{el}=T_{L}-T_{R}$) is zero and Case (b) the electron current is set up by an electronic temperature gradient ($\Delta T_{el}=T_{L}-T_{R}\neq 0$) at zero voltage. Each case operates at zero $\Delta T_{ph}$ to ensure that only electron current influences the phonon distribution. To put things simple, reservoirs $H$ and $C$ are merged to a single reservoir $B$ with temperature $T_B$ and relaxation rate $\beta$. The modified setups in Case (a) and (b) are depicted in Fig.~\ref{2a} and \ref{2b} respectively. In each operating point, the direction of phonon current is decided by the gradient between $T_B$ and dot temperature $T_M$.  \\
\indent The preceding section established that $I_{ph}^{Q}$ vanishes at $\lambda=0$. As we turn on a finite $\lambda$, the magnitude of $I_{ph}^{Q}$ increases with voltage and $\Delta T_{el}$. Both Case (a) and (b) observe this via Fig.~\ref{2c} and \ref{2d} respectively. However, we find that the direction of $I_{ph}^{Q}$ is opposite in these cases. Figure \ref{2c} notes that in Case (a), the dot phonons relax to the reservoir and the magnitude of $I_{Ph}^{Q}$ increases in the negative direction. On the other way, Fig.~\ref{2d} shows that in Case (b), the dot becomes phonon deficient and the reservoir pumps phonons into the dot. This clearly indicates that $\lambda$ acts like a phonon switch whereas voltage and $\Delta T_{el}$ control the direction of $I_{Ph}^{Q}$. These results can be utilized in the design of thermal transistors and rectifiers which aim to implement digital logic by modulating the heat current.\\
\begin{center}
	\begin{figure}[!htb]
		\subfigure[]{\includegraphics[width=0.225\textwidth, height=0.18\textwidth]{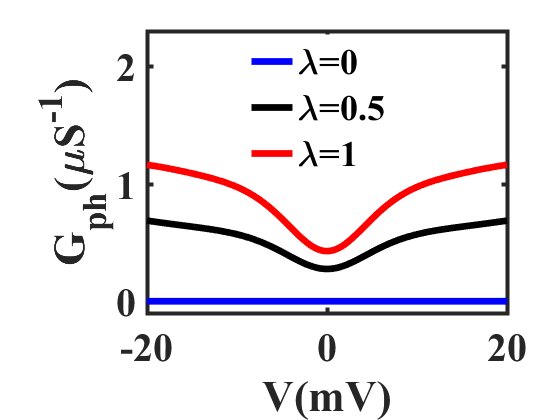}\label{3a}}
		\quad
		\subfigure[]{\includegraphics[width=0.225\textwidth, height=0.18\textwidth]{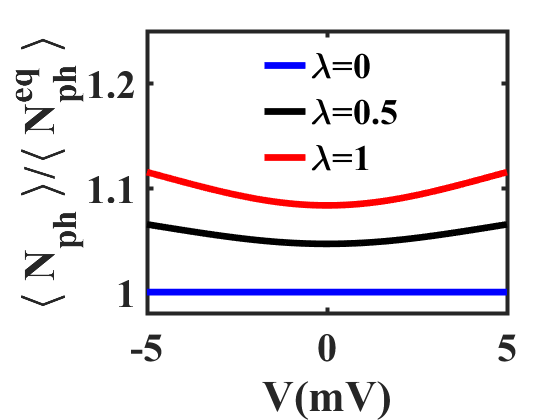}\label{3b}}
		\quad
		\subfigure[]{\includegraphics[width=0.225\textwidth, height=0.18\textwidth]{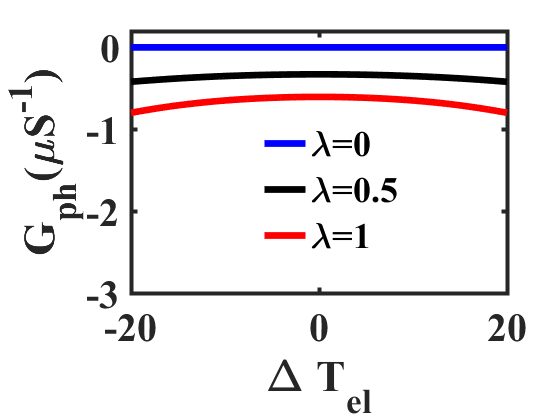}\label{3c}}
		\quad
		\subfigure[]{\includegraphics[width=0.225\textwidth, height=0.18\textwidth]{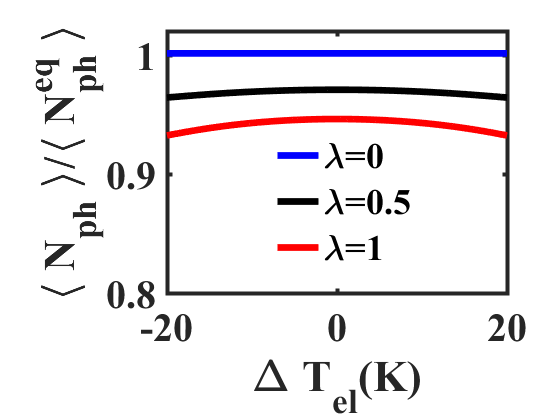}\label{3d}}
		\caption{Study of electron induced phonon generation and absorption at finite electron-phonon interaction. (a) Variation of $G_{ph}$ with voltage for different $\lambda$ in Case (a). (b) Positive deviation of $\langle N_{ph}^{eq}\rangle$ from  $\langle N_{ph}^{eq}\rangle$ as voltage increases for finite $\lambda$ in Case (a). The magnitude of phonon current increases with voltage when $\lambda$ is non-zero (c) Variation of $G_{ph}$ as a function of $\Delta T_{el}$ for different $\lambda$ in Case (b). (d) Negative deviation of $\langle N_{ph}^{eq}\rangle$ from the level of $\langle N_{ph}^{eq}\rangle$ as $\Delta T_{el}$ increases for non-zero $\lambda$ in Case (b). We infer that in Case (a) phonons accumulate in the dot and in Case (b) phonons get absorbed and explain opposite polarity of phonon current in the two cases.}
		\label{fig3}
	\end{figure}
\end{center}

\indent The anomaly in the direction of $I_{Ph}^{Q}$ can be explained by studying the phonon generation rate ($G_{ph}$) in Case (a) and (b). Figure \ref{3a} notes that in Case (a), $G_{ph}$ is positive and  rises with voltage when $\lambda$ is finite. Hence, phonons accumulate in the dot as $\langle N_{ph}\rangle$ exceeds $\langle N_{ph}^{eq}\rangle$. The voltage response of $\langle N_{ph}\rangle$ is presented in Fig.~\ref{3b}. In contrast, Fig.~\ref{3c} observes that in Case (b),  $G_{ph}$ becomes negative and falls with $\Delta T_{el}$ when $\lambda$ is non-zero. Hence the rise of $\Delta T_{el}$ facilitates phonon absorption and  $\langle N_{ph}\rangle$ falls below the level of  $\langle N_{ph}^{eq}\rangle$ as depicted in Fig.~\ref{3d}. As a result in Case(b), reservoirs pump phonons into the dot. \\

\begin{center}
	\begin{figure}
		\subfigure[]{\includegraphics[width=0.225\textwidth, height=0.18\textwidth]{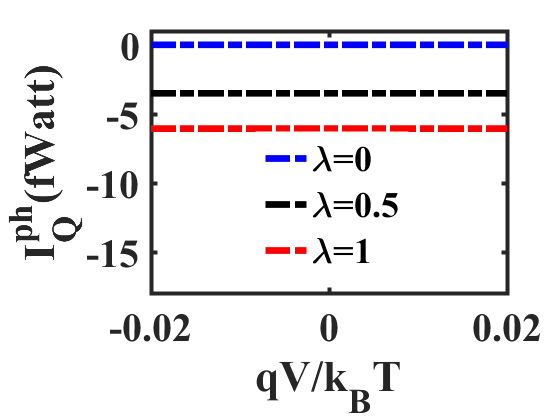}\label{4a}}
		\quad
		\subfigure[]{\includegraphics[width=0.225\textwidth, height=0.18\textwidth]{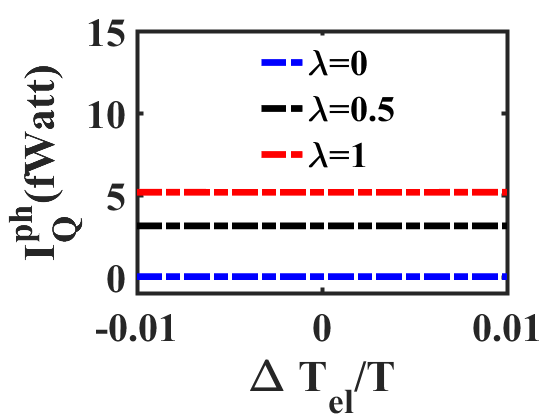}\label{4b}}
		\quad
		\subfigure[]{\includegraphics[width=0.225\textwidth, height=0.18\textwidth]{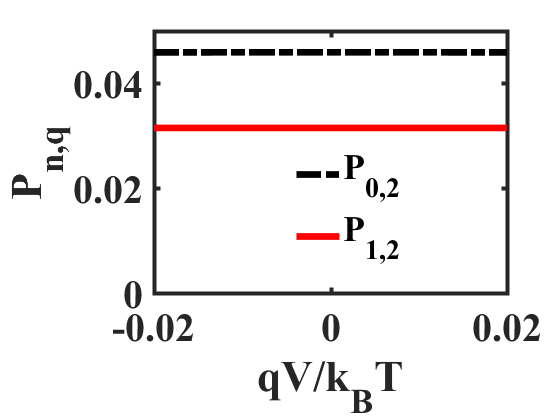}\label{4c}} 
		\quad
		\subfigure[]{\includegraphics[width=0.225\textwidth, height=0.18\textwidth]{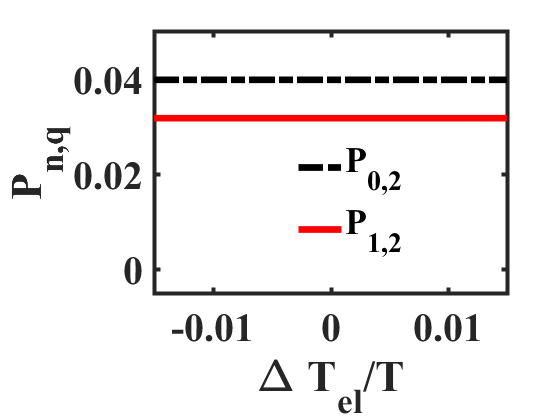}\label{4d}}
		\caption{ Linear response of $I_Q^{ph}$. (a) Variation of $I_Q^{ph}$ with voltage in the linear regime for different $\lambda$ in Case (a). (b) Variation of $I_Q^{ph}$ with $\Delta T_{el}$ in the linear regime for different $\lambda$ in Case (b). (c) Variation of $P_{n,q}$ with linear voltage in Case (a) at non-zero $\lambda$ ($\lambda=0.5$).  (d) Variation of $P_{n,q}$ with linear range of electronic temperature bias in Case (b) at non-zero $\lambda$ ($\lambda=0.5$). This figure notes that in the linear regime the probabilities of the states $|n,q\rangle$ remain constant and consequently $I_Q^{ph}$ show no variation with voltage or $\Delta T_{el}$.}
		\label{fig4}
	\end{figure}
\end{center}

\indent One should note that the variation of $I_{ph}^{Q}$  with voltage  and $\Delta T_{el}$ takes place in a non-linear fashion. Now it is essential to test the features of $I_{ph}^{Q}$ in the linear regime. From the fundamental standpoint, this is imperative since Onsager's reciprocity should be validated in the linear regime. Figure \ref{4a} depicts that in Case (a), the phonon heat current does not vary with voltage in the linear limit (q$V<<k_{B}T$). Similarly, Fig.\ref{4b} shows that in Case (b), $I_{Ph}^{Q}$ does not change with $\Delta T_{el}$ in the linear range ($\Delta T_{el}<<T$). To explore the linear response of $I_Q^{ph}$, we test the nature of many body electron-phonon probabilities $P_{n,q}$ of the states $|n,q\rangle$ in Case (a) and (b), when  $\lambda$ is non-zero. Figure.\ref{4c} presents that in Case (a), $P_{n,q}$ does not vary with voltage. Similarly, Fig.\ref{4d} depicts that in Case (b), $P_{n,q}$ remains constant with $\Delta T_{el}$. Hence, in the linear regime, the average population of phonons in the dot ($\langle N_{ph}\rangle=\sum\limits_{n,q}^{}qP_{n,q}$) remains unaltered and $I_{Ph}^{Q}$ remains constant as dictated in \eqref{Eq15}. This implies that the electron and phonon transport are uncoupled in the linear regime .\\
\begin{center}
	\begin{figure}
		\subfigure[]{\includegraphics[width=0.225\textwidth, height=0.18\textwidth]{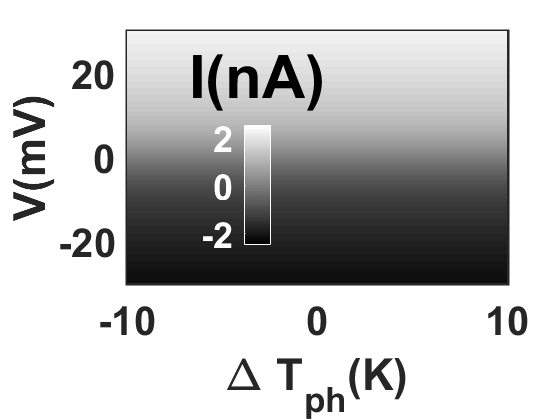}\label{5a}}
		\quad
		\subfigure[]{\includegraphics[width=0.225\textwidth, height=0.18\textwidth]{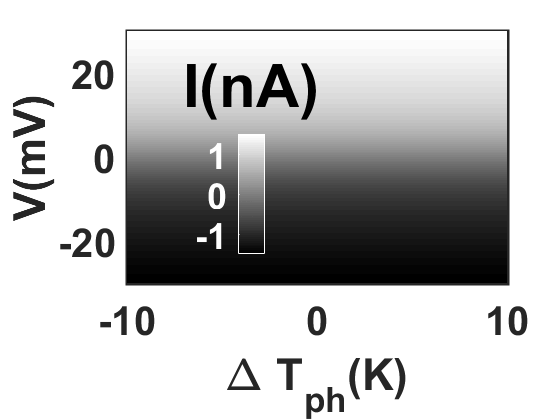}\label{5b}}
		\quad
		\subfigure[]{\includegraphics[width=0.225\textwidth, height=0.18\textwidth]{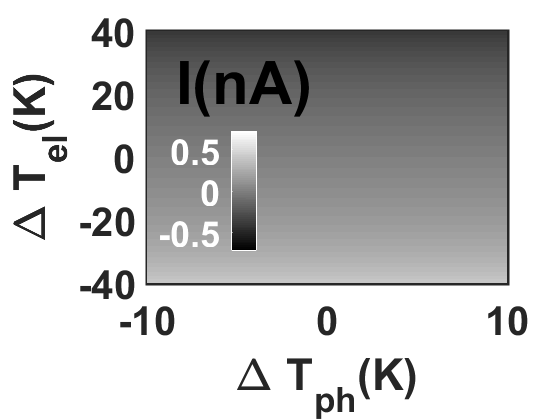}\label{5c}}
		\quad
		\subfigure[]{\includegraphics[width=0.225\textwidth, height=0.18\textwidth]{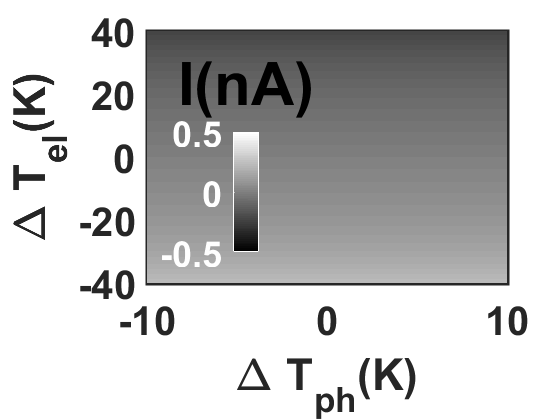}\label{5d}}
		
		\caption{Test of the inter-dependence of $I$ and $\Delta T_{ph}$. Color variation of $I$ as a function of $\Delta T_{ph}$ and $V$ for (a) $\lambda=0$ and (b) $\lambda=0.5$ when $\Delta T_{el}=0$. Color variation of $I$ as a function of $\Delta T_{ph}$ and $\Delta T_{el}$ for (c) $\lambda=0$ and (d) $\lambda=0.5$ at zero voltage. It implies that in our setup $I$ can not be stimulated by $\Delta T_{ph}$ even in the non-linear regime.}
		\label{fig5}
	\end{figure}
\end{center}
\indent After the discussion of linear response of $I_Q^{ph}$, it is essential to examine the inter-relation of electron current $I$ and the phonon temperature gradient ($\Delta T_{ph}=T_H-T_C$). In this course, first we set $\Delta T_{el}$ at zero and test the variation of $I$ as a function of voltage and $\Delta T_{ph}$. Figure.\ref{5a} and \ref{5b} depict the color variation of $I$ as a function of $V$ and $\Delta T_{ph}$ for $\lambda=0$ and $\lambda=0.5$ respectively. It is observed that even in the non-linear regime, $I$ shows no dependence on $\Delta T_{ph}$. Next, we study the variation of $I$ with $\Delta T_{el}$ and $\Delta T_{ph}$ at zero voltage. The color variation of $I$ as a function of $\Delta T_{el}$ and $\Delta T_{ph}$  is depicted in Fig.\ref{5c} and \ref{5d} for $\lambda=0$ and $\lambda=0.5$ respectively. Here also, we notice that $I$ shows no dependence on $\Delta T_{ph}$. This result is consistent with the previous report\cite{Jing} which demonstrated why $I$ can not be stimulated by $\Delta T_{ph}$ in quantum dot based devices with a single phonon mode. In the linear regime, several charge and heat currents $I$,$I_Q^{el}$ and $I_Q^{ph}$ are related to voltage and various temperature gradient $V$, $\Delta T_{el}$ and $\Delta T_{ph}$ by the Onsager's matrix, such that
\begin{equation*}
\begin{bmatrix}
{\frac{\mathlarger{I}}{\mathlarger{q}}} \\ \\
{\frac{\mathlarger{I_Q^{el}}}{\mathlarger{k_B T}}} \\ \\
{\frac{\mathlarger{I_Q^{ph}}}{\mathlarger{k_B T}}}
\end{bmatrix}=
\begin{bmatrix}
L_{11} & L_{12} & L_{13}\\ \\
L_{21} & L_{22} & L_{23}\\ \\
L_{31} & L_{32} & L_{33}
\end{bmatrix}
\begin{bmatrix}
qV \\ \\
k_{B}\Delta T_{el}\\ \\
k_{B}\Delta T_{ph} 
\end{bmatrix}.
\end{equation*}
\indent In the present scenario, we find that $L_{13}=L_{31}=0$ and $L_{23}=L_{32}=0$. It confirms that Onsager's reciprocity is obeyed. Now the dependence of $I_Q^{ph}$ on the voltage and $\Delta T_{el}$, immediately hints at the concept of a novel thermal conductivity coefficient which is different fron the conventional electronic thermal conductivity. We will elaborate on it in the next subsection.
\begin{center}
	\begin{figure}
		\subfigure[]{\includegraphics[width=0.225\textwidth, height=0.18\textwidth]{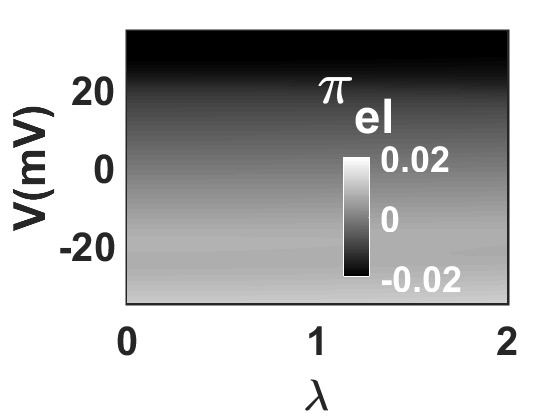}\label{6a}}
		\quad
		\subfigure[]{\includegraphics[width=0.225\textwidth, height=0.18\textwidth]{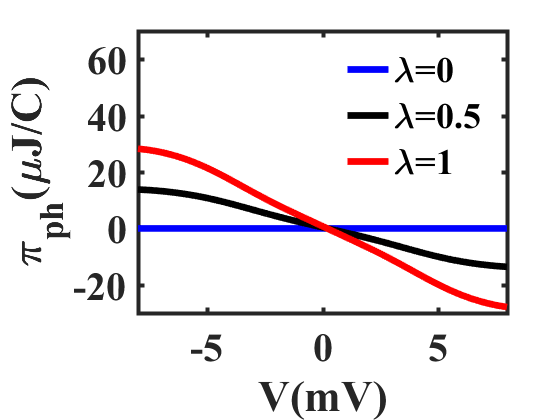}\label{6b}}
		\quad
		\subfigure[]{\includegraphics[width=0.225\textwidth, height=0.18\textwidth]{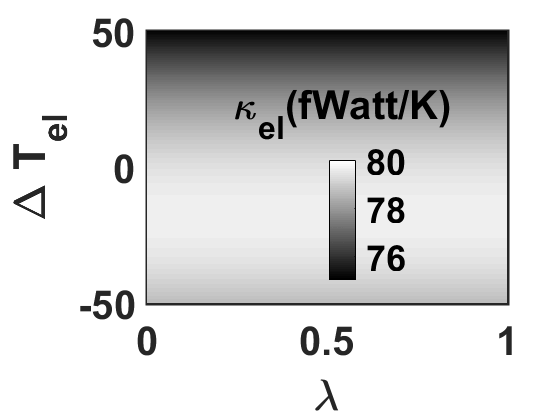}\label{6c}}
		\quad
		\subfigure[]{\includegraphics[width=0.225\textwidth, height=0.18\textwidth]{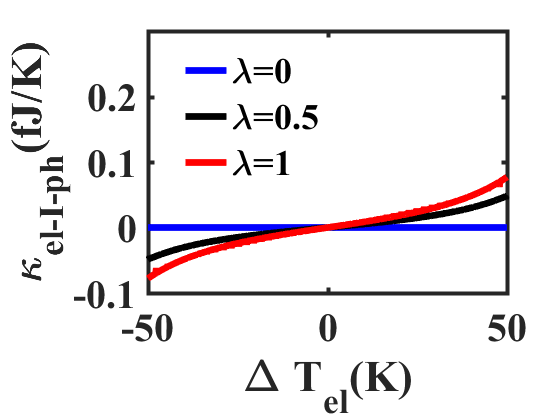}\label{6d}}
		\caption{Phonon Peltier and electron-assisted phonon thermal conductivity coefficients. (a) Color variation of electronic Peltier coefficient $\pi_{el}$ as a function of $\lambda$ and $V$. (b) Variation of the phonon Peltier coefficient as a function of voltage for different $\lambda$. (c) Color variation of electronic thermal conductivity $\kappa_{el}$ as a function of $\lambda$ and $\Delta T_{el}$. (d) Variation of $\kappa_{el-I_ph}$ as a function of $\Delta T_{el}$ for different $\lambda$. We infer from this figure that $\pi_{el}$ and $\kappa_{el}$ are independent of $\lambda$ although $\pi_{ph}$ and $\kappa_{el-I_ph}$ are strong functions of $\lambda$.}
		\label{fig6}
	\end{figure}
\end{center}
\subsection{Non-linear phonon Peltier and electron-induced phonon thermal conductivity}
\indent The conventional electronic Peltier coefficient is defined as $\pi_{el}=\frac{dI_Q^{el}}{dI}$, when there is no electronic temperature gradient ($\Delta T_{el}=0$). In analogy with the electronic Peltier coefficient, our previous work \cite{nsr} has introduced the phonon Peltier coefficient as $\pi_{ph}=\frac{dI_Q^{ph}}{dI}$. In this subsection, we draw a comparative study between the two Peltier co-efficients.\\
\indent  Both Peltier coefficients are computed in the differential form to avoid the singularity at the short-circuit point ($V=0$). First, we plot the color variation of $\pi_{el}$ as a function of voltage and $\lambda$ in Fig.\ref{6a}. It shows that $\pi_{el}$ varies with voltage but remains completely independent of $\lambda$. It implies that in the absence of an electronic temperature gradient, the ratio of the electronic heat current $I_Q^{el}$ and the charge current does not vary $\lambda$. In contrast, Fig.\ref{6b} notes that  $\pi_{ph}$ is a joint function of voltage and $\lambda$. $\pi_{ph}$ vanishes when $\lambda$ is zero and it varies linearly with the voltage as we switch on a finite $\lambda$. Earlier Fig.\ref{2a} showed that the magnitude of $I_Q^{ph}$ increases with $\lambda$ and Fig.\ref{fig5} showed that the magnitude of $I$ falls with $\lambda$ at fixed voltage. Therefore, the absolute magnitude of $\pi_{ph}$ increases with $\lambda$ as shown in Fig.\ref{6b}. Since $I_Q^{ph}$ varies with voltage only in the non-linear regime, we mention the phonon Peltier co-efficient ($\pi_{ph}$) as \textit{non-linear phonon Peltier coefficient}.\\
 \indent The concept of $\pi_{ph}$ was manifested since voltage bias stimulates and modulates $I_Q^{ph}$, when $\Delta T_{el}$ is zero. We have already observed that, $I_Q^{ph}$ is a joint function of $V$ and $\Delta T_{el}$. This motivates us to propose a new thermal conductivity coefficient $\kappa_{el-I-ph}$, which is different from the conventional electronic thermal conductivity $\kappa_{el}$. Both thermal conductivity coefficients $\kappa_{el}$ and $\kappa_{el-I-ph}$ are defined as,
\[
\left. \kappa_{el}=-\frac{\partial I_Q^{el}}{\partial \Delta T_{el}} \right|_{\mathrlap{I=0}},
\]
\[
\left. \kappa_{el-I-ph}=-\frac{\partial I_Q^{ph}}{\partial \Delta T_{el}} \right|_{\mathrlap{I=0}}.
\]
The color plot shown in Fig.\ref{6c}, depicts that $\kappa_{el}$ remains constant with $\lambda$ and shows minor variation with $\Delta T_{el}$. On the other hand, Fig.\ref{6d} notes that the absolute magnitude of $\kappa_{el-I-ph}$ increases with both $\lambda$ and $\Delta T_{el}$. At constant $\lambda$,  the open-circuit voltage of the device increases with the increase of $\Delta T_{el}$. On the other hand, the magnitude of $I_Q^{ph}$ increases with the voltage bias as shown in Fig.\ref{2a}. Therefore, the absolute magnitude of $\kappa_{el-I-ph}$ rises with $\Delta T_{el}$ as shown in Fig.\ref{6d}. Therefore, the absolute magnitude of $\kappa_{el-I-ph}$ rises with $\Delta T_{el}$. Since $I_Q^{ph}$ varies with voltage and $\Delta T_{el}$ only in the non-linear regime, we mention $\kappa_{el-I-ph}$, as the \textit{non-linear electron assisted phonon thermal conductivity}.
\begin{center}
	\begin{figure}
		\subfigure[]{\includegraphics[width=0.225\textwidth, height=0.18\textwidth]{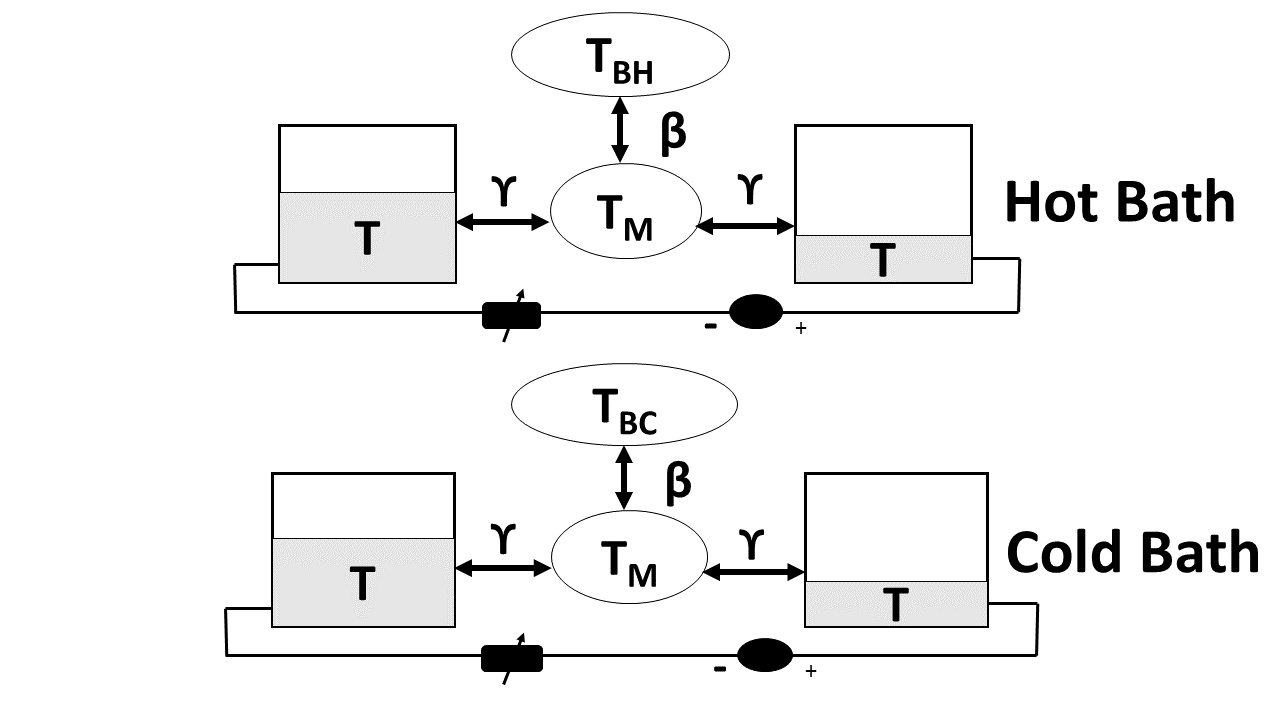}\label{7a}}
		\quad
		\subfigure[]{\includegraphics[width=0.225\textwidth, height=0.18\textwidth]{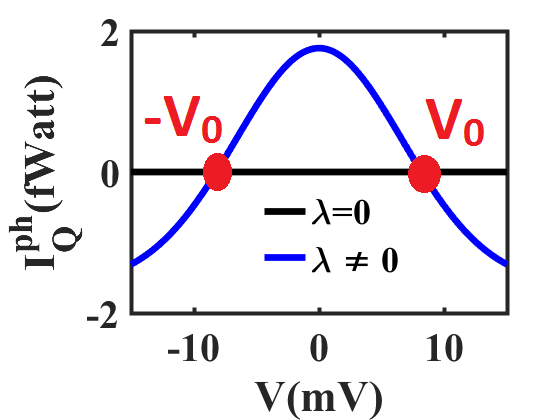}\label{7b}}
		\quad
		\subfigure[]{\includegraphics[width=0.225\textwidth, height=0.18\textwidth]{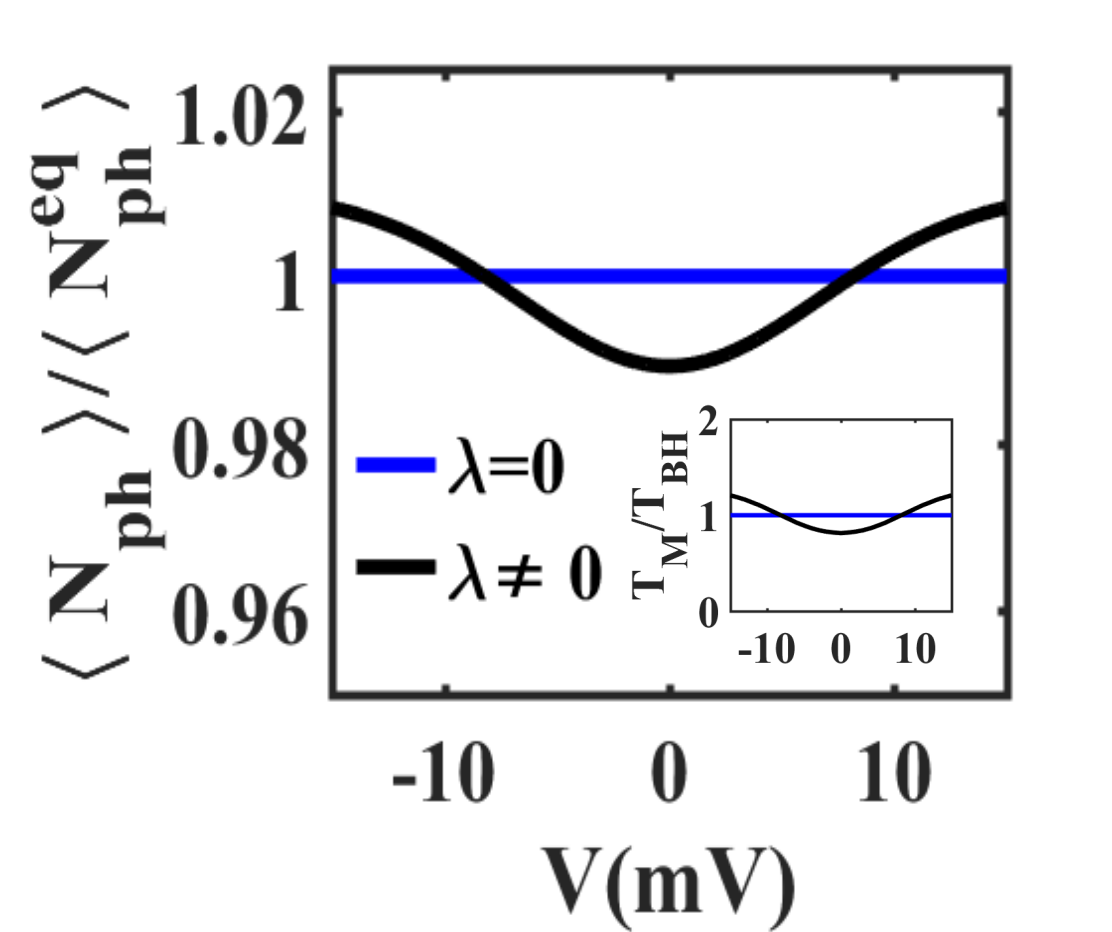}\label{7c}}
		\quad
		\subfigure[]{\includegraphics[width=0.225\textwidth, height=0.18\textwidth]{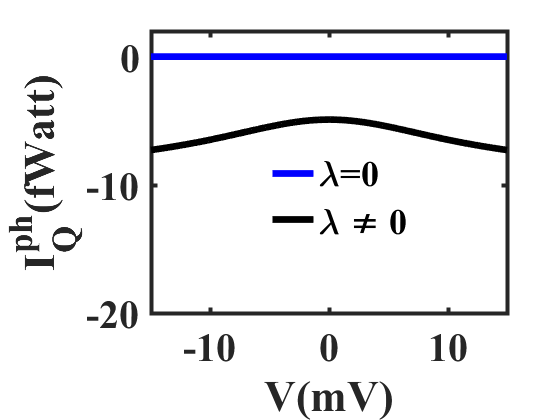}\label{7d}}
		\quad
		\subfigure[]{\includegraphics[width=0.225\textwidth, height=0.18\textwidth]{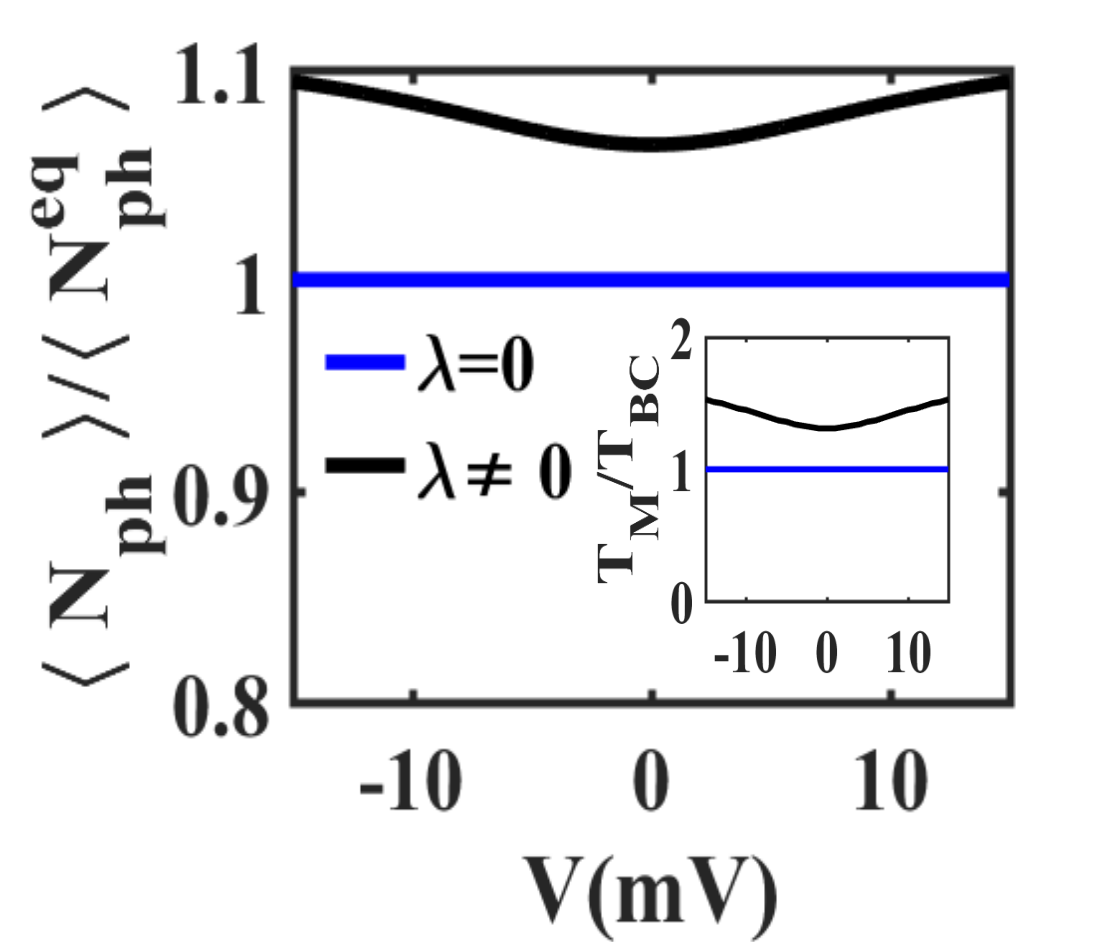}\label{7e}}
		\quad
		\subfigure[]{\includegraphics[width=0.225\textwidth, height=0.18\textwidth]{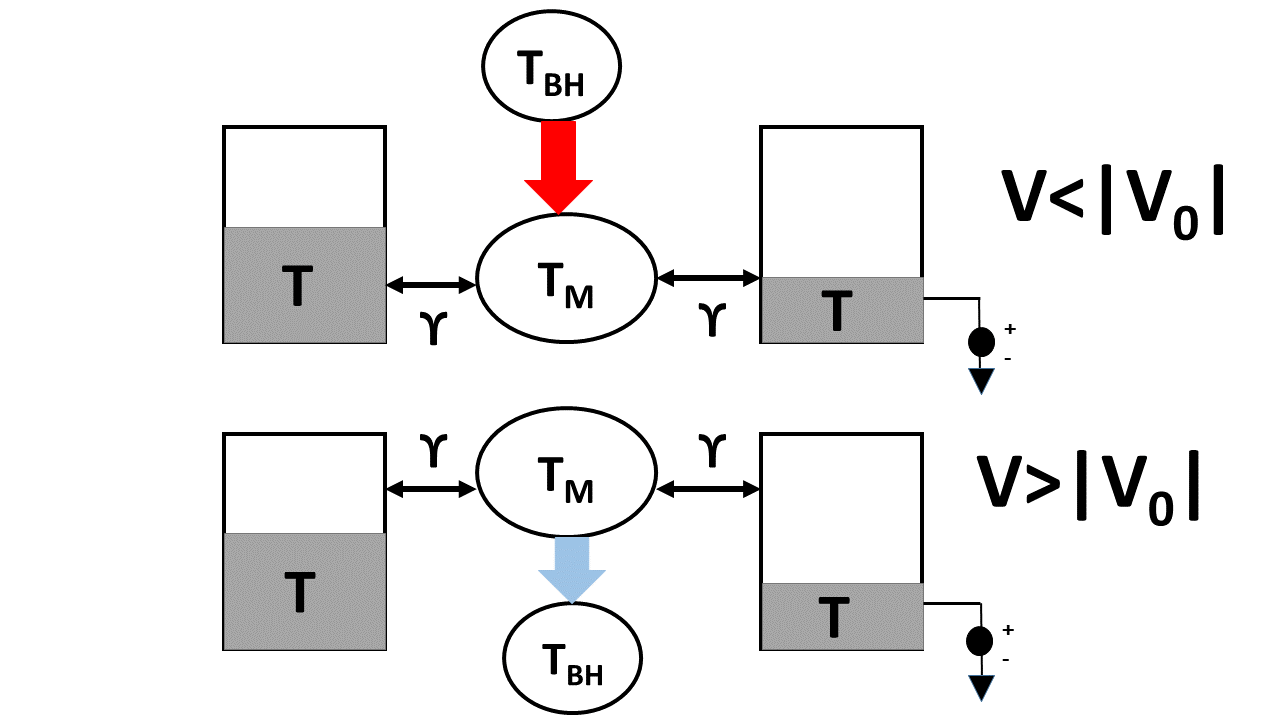}\label{7f}}
		\caption{Control of $I_Q^{ph}$ by voltage bias in Case (a), when $\Delta T_{el}=0$. Two situations are considered: Case (HR) when the reservoir is hotter than the electronic contacts ($T_L=T_R=100$ K, $T_B=T_{BH}=150$ K) and  Case (CR) when the reservoir is colder than the electronic contacts ($T_L=T_R=150$ K, $T_B=T_{BC}=100$ K). (a) The modified device diagrams in the two cases. (b) Voltage Variation of $I_Q^{ph}$ for $\lambda=0$ and $\lambda \neq 0$ in Case (HR). (c) Voltage variation of $\langle N_{ph}\rangle$ in Case (HR). Inset shows the variation of dot temperature $T_M$ with voltage. (d) Voltage variation of $I_Q^{ph}$ for $\lambda=0$ and $\lambda \neq 0$ in Case (CR). (e) Voltage variation of $\langle N_{ph}\rangle$ with voltage in Case (CR). Inset shows the variation of $T_M$ with voltage. (f) The schematic of $I_Q^{ph}$ for Case (HR). This figure shows that in Case (a), the polarity of $I_Q^{PH}$ can be reversed by driving the voltage in Case (HR).}
		\label{fig7}
	\end{figure}
\end{center}	
\subsection{Engineering of $I_Q^{ph}$ and $I_Q^{el}$ by varying reservoir temperature}
\indent It is evident from \eqref{Eq7},\eqref{Eq8} and \eqref{Eq12} that $I_Q^{ph}$ can be controlled by differing the temperature ($T_{B}$) of the reservoir from the  temperature of the electronic contacts ($T_{L(R)}$). In this subsection, we demonstrate the engineering  of $I_Q^{ph}$  in two situations: Case (HR) where, the reservoir is hotter than the contacts ($T_{B}=T_{BH}>T_{L(R)}$) and Case (CR) where, the reservoir is colder than the contacts ($T_{B}=T_{BC}<T_{L(R)}$). First, we illustrate these cases in the limit of Case (a) where $I_Q^{ph}$ is controlled by driving the voltage bias at zero electronic temperature gradient ($\Delta T_{el}=0$) .\\

\begin{center}
	\begin{figure}
		\subfigure[]{\includegraphics[width=0.225\textwidth, height=0.18\textwidth]{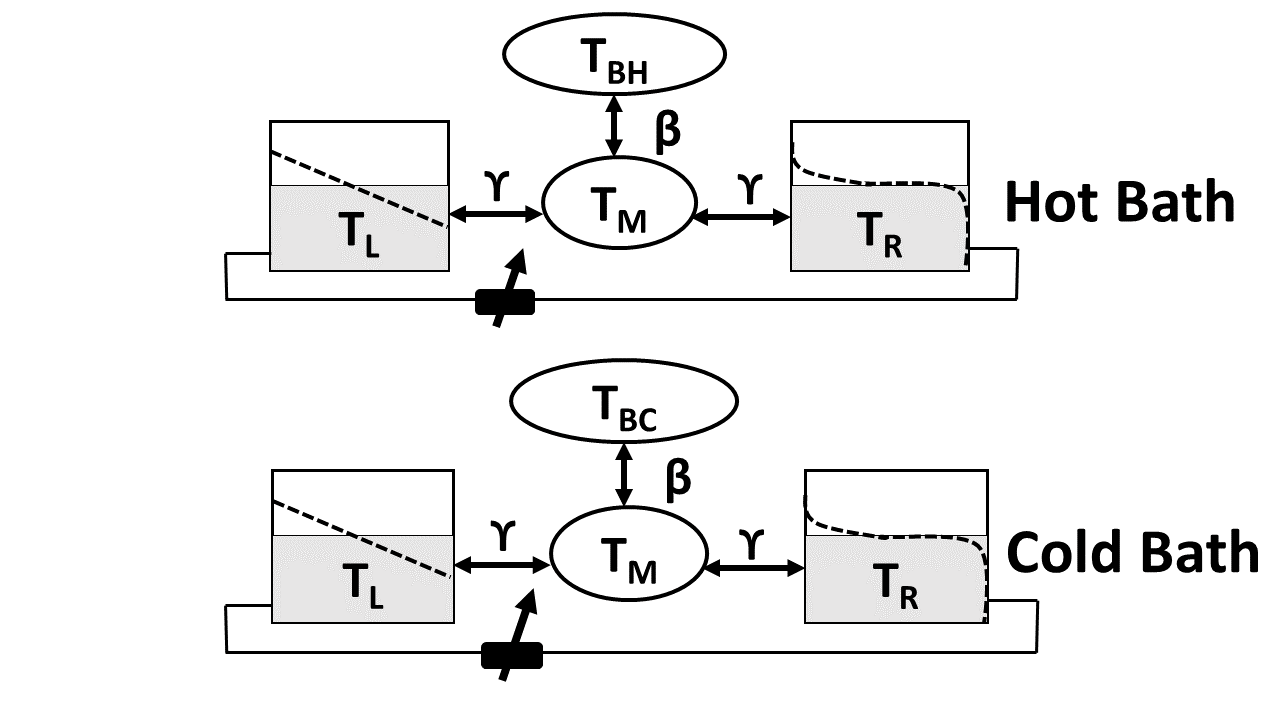}\label{8a}}
		\quad
		\subfigure[]{\includegraphics[width=0.225\textwidth, height=0.18\textwidth]{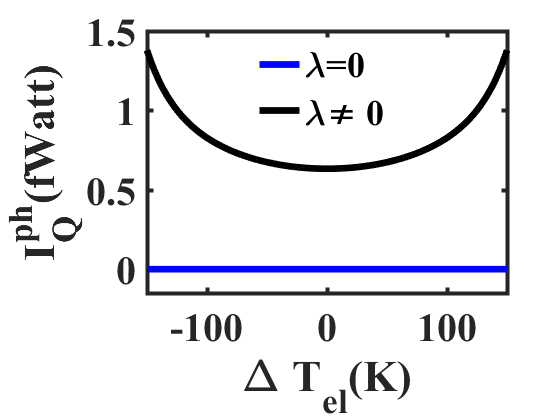}\label{8b}}
		\quad
		\subfigure[]{\includegraphics[width=0.225\textwidth, height=0.18\textwidth]{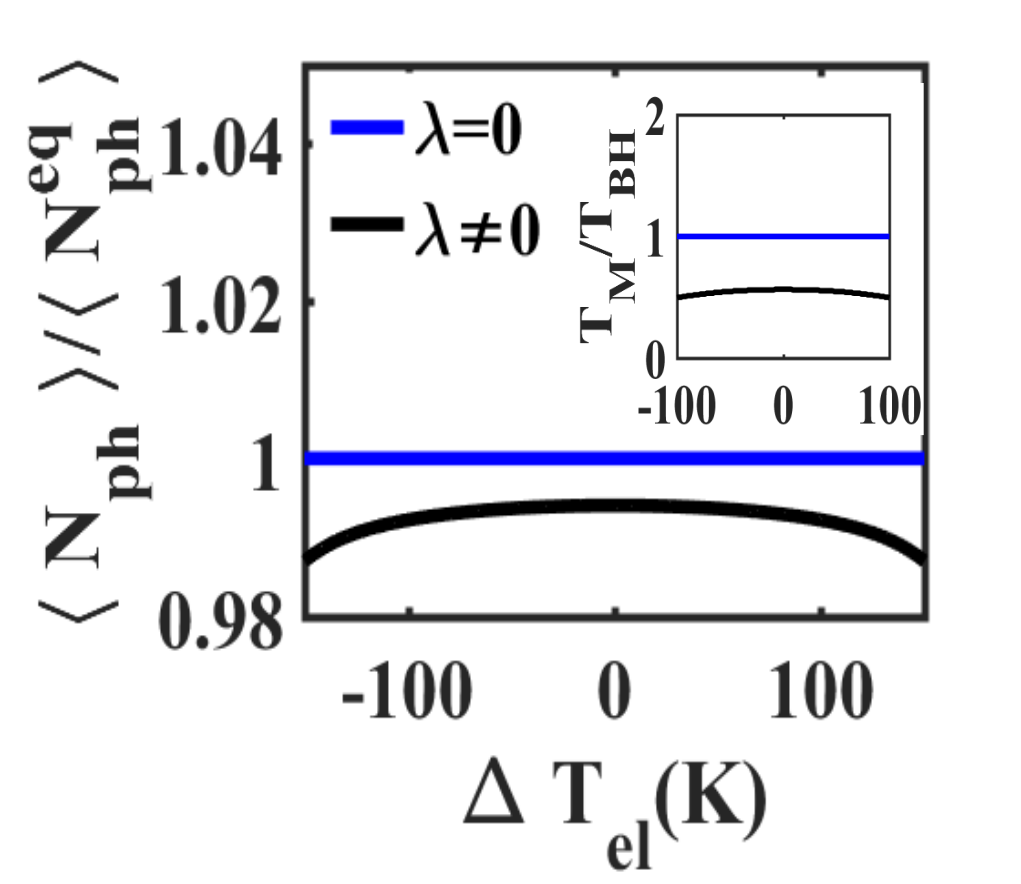}\label{8c}}
		\quad
		\subfigure[]{\includegraphics[width=0.225\textwidth, height=0.18\textwidth]{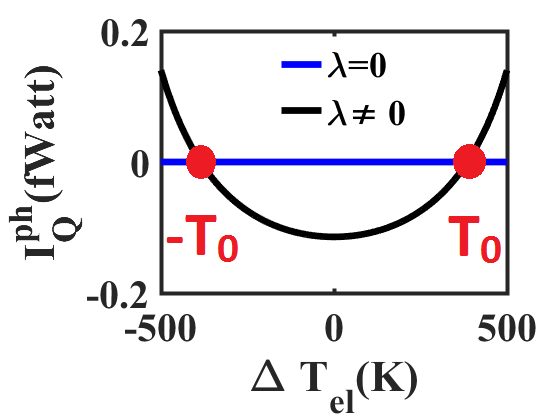}\label{8d}}
		\quad
		\subfigure[]{\includegraphics[width=0.225\textwidth, height=0.18\textwidth]{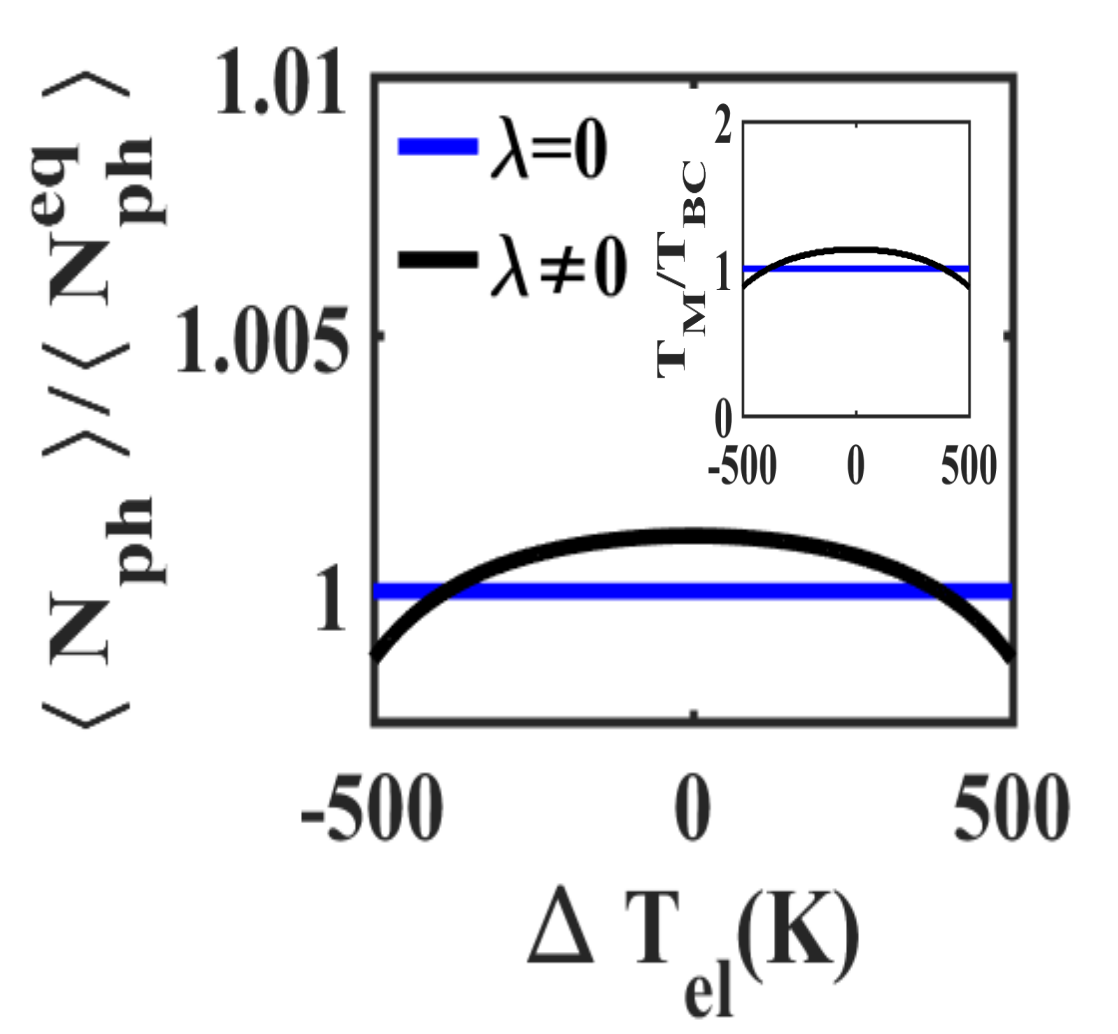}\label{8e}}
		\quad
		\subfigure[]{\includegraphics[width=0.225\textwidth, height=0.18\textwidth]{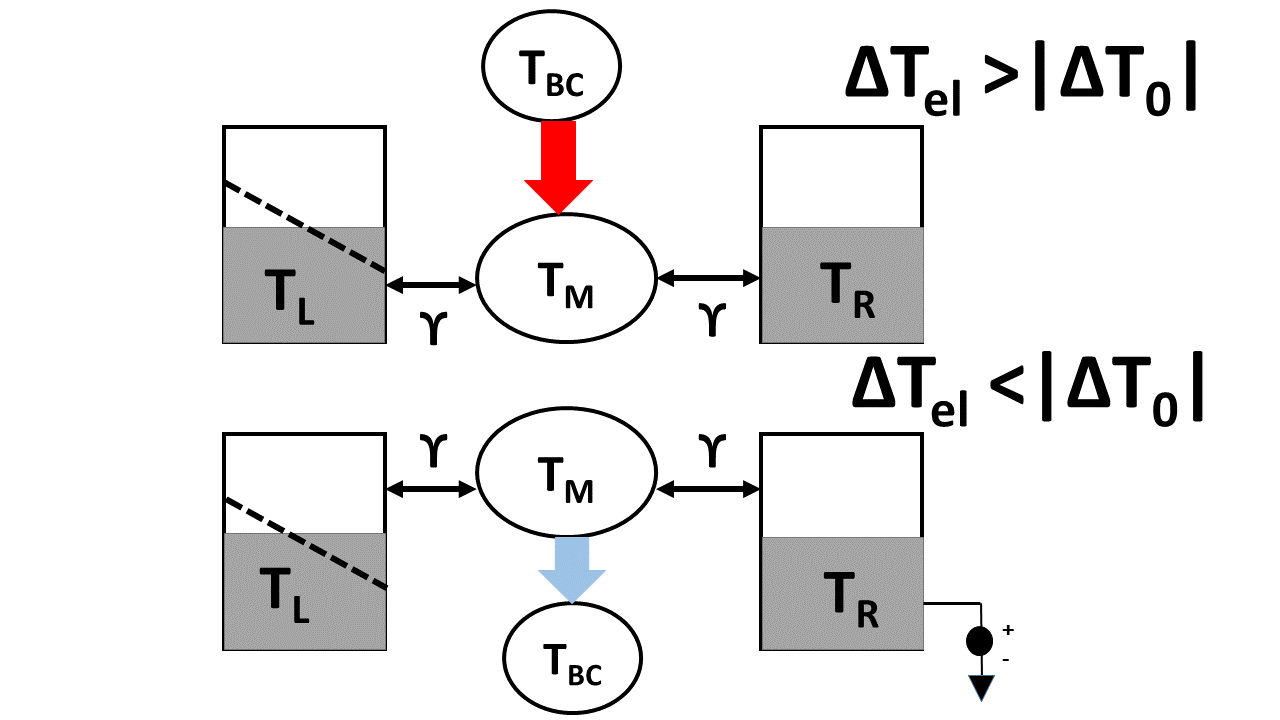}\label{8f}}
		\caption{Control of $I_Q^{ph}$ by driving $\Delta T_{el}$, at $V=0$. Two cases are considered: Case (HR) Case (HR) when the reservoir is hotter than the electronic contacts ($T_H=T_C=T_{BH}=300 K>T_{L}=T_{R}=100 K$) and Case (CR) Cold Bath: when the reservoir is colder than the electronic contacts ($T_H=T_C=T_{BC}=250 K<T_{L}=T_{R}=350 K$). (a) The modified device schematics in two cases. (b) Variation of $I_Q^{ph}$ as a function of $\Delta T_{el}$ for $\lambda=0$ and $\lambda \neq 0$ for Case (HR). (c) Variation of $\langle N_{ph}\rangle$ with $\Delta T_{el}$ in Case (HR). Inset shows the variation of $T_M$ with $\Delta T_{el}$ in the same case. (d) Variation of $I_Q^{ph}$ with $\Delta T_{el}$ for $\lambda=0$ and $\lambda \neq 0$. (e) Variation of $\langle N_{ph}\rangle$ with $\Delta T_{el}$. Inset shows the variation of $T_M$ with $\Delta T_{el}$ in the same case. (f) The schematic of $I_Q^{ph}$ in Case (CR). This figure infers that the schematic of that in Case (b), the polarity odf $I_Q^{ph}$ can be flipped by driving $\Delta T_{el}$ in Case (CR).}
		\label{fig8}
	\end{figure}
\end{center}

\indent We present the setups in Case (HR) and (CR) in Fig.\ref{7a}. Figure \ref{7b} plots the voltage response of $I_Q^{ph}$ in Case (HR). In this case,  $I_Q^{ph}$ flows from the dot to the reservoir in the low voltage range but reverses it's polarity at $|V|=V_0$, when $\lambda$ is non-zero. This non-trivial phenomenon can be explained by studying the voltage variation of $\langle N_{ph}^{eq}\rangle$ depicted in Fig.\ref{7c}. As we increase the voltage, phonons accumulate in the dot and at $V=|V_0|$, the average phonon number in the dot exceeds the equilibrium phonon number $\langle N_{ph}^{eq}\rangle$ of the reservoir. Hence, the polarity of $I_Q^{ph}$ is flipped. However, Fig.\ref{7d} captures no such non-trivial polarity reversal of $I_Q^{ph}$ in Case (CR). In this case, Fig.\ref{7e} notes that  $\langle N_{ph}\rangle$ is always higher than the mark of $\langle N_{ph}^{eq}\rangle$ and $I_Q^{ph}$ is directed from the dot to the cold reservoir. The voltage variation of $\langle N_{ph}\rangle$ presented in Fig.\ref{7c} and \ref{7e} is consistent with the voltage variation of the dot temperature $T_M$ shown in their respective insets. It is interesting to observe that the electron current emanating from electronic contacts can pump phonons into the hotter reservoir.  This is a type of counter-intuitive heating of a hot reservoir. Figure \ref{7f} depicts the schematic of counter-intuitive heating in Case (HR).
\begin{center}
	\begin{figure}
		\subfigure[]{\includegraphics[width=0.225\textwidth, height=0.18\textwidth]{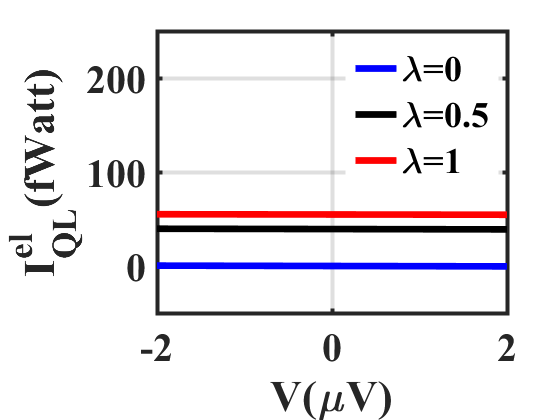}\label{9a}}
		\quad
		\subfigure[]{\includegraphics[width=0.225\textwidth, height=0.18\textwidth]{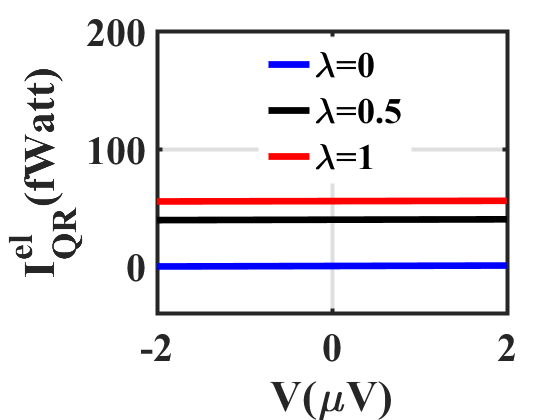}\label{9b}}
		\quad
		\subfigure[]{\includegraphics[width=0.225\textwidth, height=0.18\textwidth]{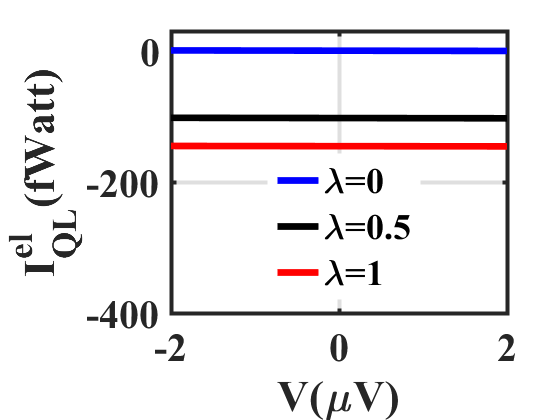}\label{9c}}
		\quad
		\subfigure[]{\includegraphics[width=0.225\textwidth, height=0.18\textwidth]{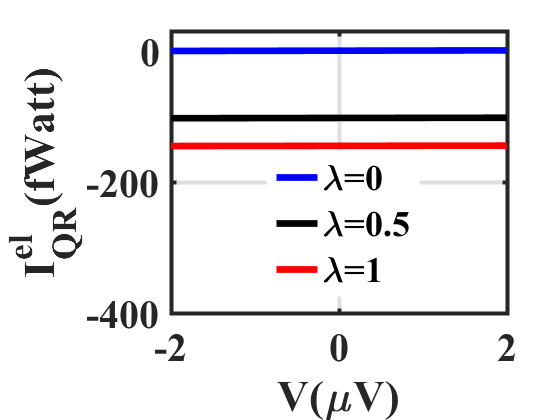}\label{9d}}
		\quad
		\subfigure[]{\includegraphics[width=0.225\textwidth, height=0.18\textwidth]{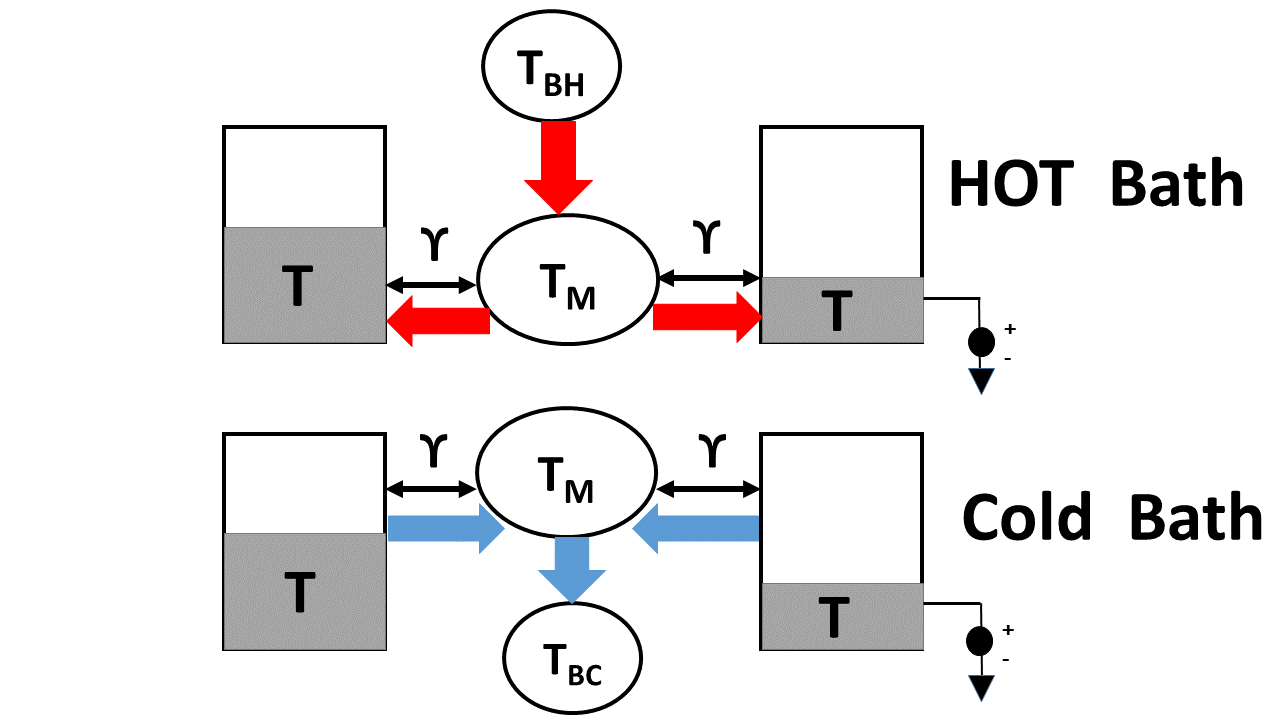}\label{9e}}
		\quad
		\subfigure[]{\includegraphics[width=0.225\textwidth, height=0.18\textwidth]{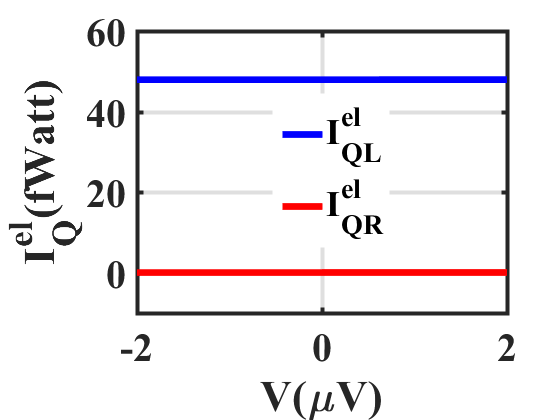}\label{9f}}
		\quad
		\subfigure[]{\includegraphics[width=0.225\textwidth, height=0.18\textwidth]{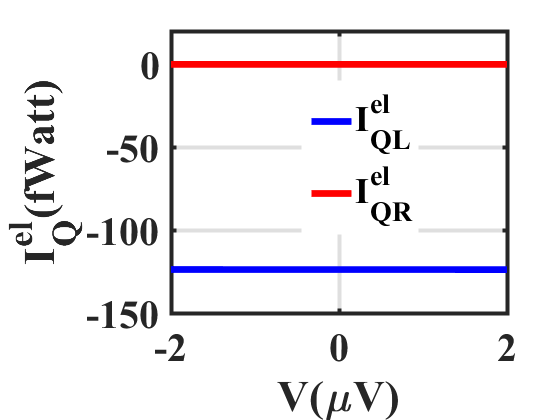}\label{9g}}
		\quad
		\subfigure[]{\includegraphics[width=0.225\textwidth, height=0.18\textwidth]{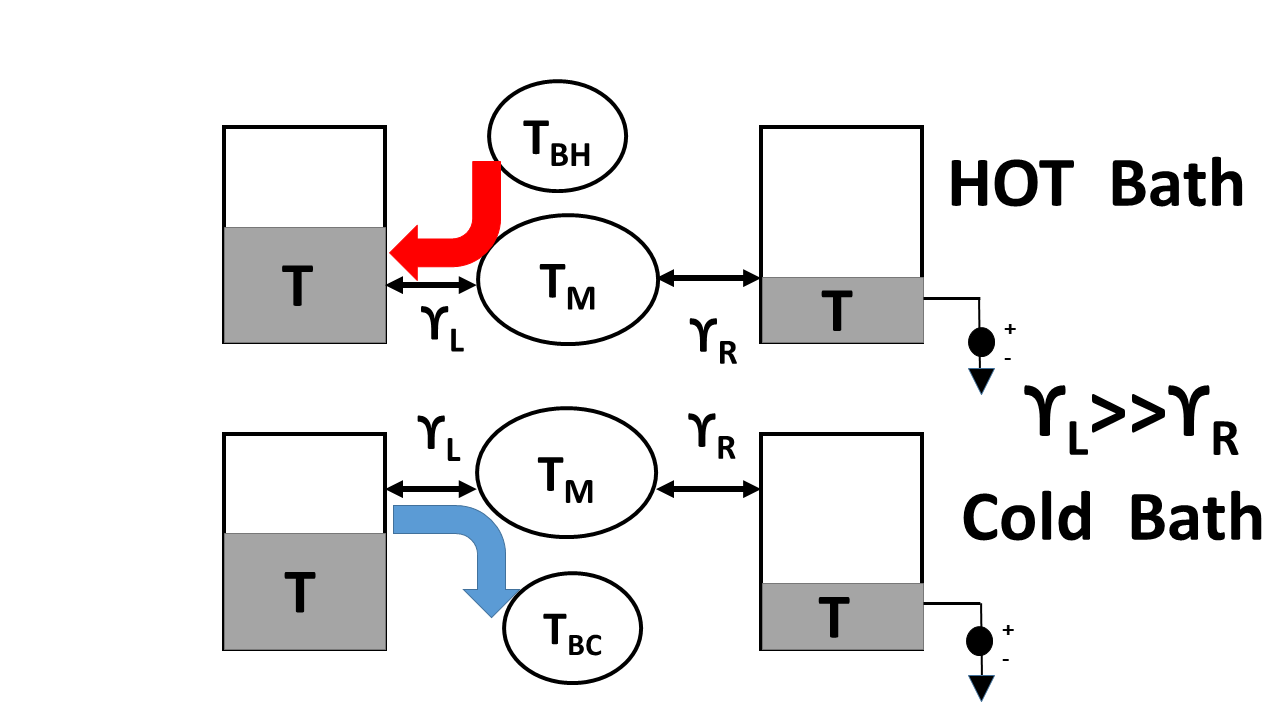}\label{9h}}
		\caption{Study of small voltage response of electronic heat current $I_{QL(R)}^{el}$ associated with contact $L(R)$, when $\Delta T_{el}=0$. We consider two cases:(1) Hot Bath: $T_H=T_C>T_L=T_R$ and (2) Cold Bath: $T_H=T_C<T_L=T_R$. (a) and (b) plot the respective voltage variation of $I_{QL}^{el}$ and $I_{QR}^{el}$ for Hot Bath as $\lambda$ varies. Similarly (c) and (d) show the respective voltage variation of $I_{QL}^{el}$ and $I_{QR}^{el}$ for Cold Bath as $\lambda$ varies. (e) Present the schematic of $I_{QL(R)}^{el}$ in two cases. (f) and(g) depict the voltage dependence of $I_{QL}^{el}, I_{QR}^{el}$ for non-zero $\lambda$ for cases (1) and (2) respectively, when $\gamma_L>>\gamma_R$. (h) presents the schematic for two cases given $\gamma_L>>\gamma_R$. }
		\label{fig9}
	\end{figure}
\end{center}

\indent An opposite response of $I_Q^{ph}$ is observed when we analyze Case (HR) and (CR) in the limit of Case (b) where we drive the electronic temperature gradient at zero voltage. Figure \ref{8a} depicts the modified device schematics in Case (HR) and (CR). Figure.\ref{8b} plots the variation of $I_Q^{ph}$ with $\Delta T_{el}$ in the Case (HR) for $\lambda=0$ and  $\lambda\neq 0$.  In this case, we note no non-triviality in the direction of $I_Q^{ph}$ and reservoir pumps phonons into the dot for all values of $\Delta T_{el}$. Fig.\ref{8c} shows that in this case, $\langle N_{ph}\rangle$ is always lower than the level of $\langle N_{ph}^{eq}\rangle$ and $I_Q^{ph}$ does not flip its polarity. In contrast, Fig.\ref{8d} depicts that in Case (CR), $I_Q^{ph}$ reverses it's direction at $\Delta T_{el}=|T_0|$ for non-zero $\lambda$ and cold reservoir pumps phonons into the dot. As explained earlier, the increase of $\Delta T_{el}$ enhances the absorption of phonons and the average phonon number in the dot falls below the level of at $\Delta T_{el}=|T_0|$. This phenomenon is explained with the help of Fig.\ref{8e}.The $\Delta T_{el}$ variation of $\langle N_{ph}\rangle$ presented in Fig.\ref{8c} and \ref{8e} is consistent with the $\Delta T_{el}$ variation of the dot temperature $T_M$ shown in their respective insets. Therefore, this case observes a counter-intuitive cooling is of a cold reservoir. Figure \ref{8f} depicts the schematic of counter-intuitive cooling in the Case (CR).\\
\indent Finally, we elaborate on the engineering of the electronic heat current ($I_Q^{el}$) in Case (HR) and (CR). In this context we rename $I_Q^{el}$ associated with the contacts $L$ and $R$ as $I_{QL}^{el}$ and $I_{QR}^{el}$ respectively. Figure \ref{9a} and \ref{9b} depict the voltage variation of $I_{QL}^{el}$ and $I_{QR}^{el}$ for different values of $\lambda$ in Case (HR). We observe that  both $I_{QL}^{el}$ and $I_{QR}^{el}$ vanishes when $\lambda=0$. When $\lambda$ is non-zero, both $I_{QL}^{el}$ and $I_{QR}^{el}$ becomes finite and they disperse uniformly into the contacts even at the short-circuit point ($V=0$). At $\lambda\neq0$, the Hot Bath push phonons into the dot and the excess heat flows away into the contacts in the form of $I_Q^{el}$. Similarly Fig.\ref{9c} and \ref{9d} explain the same phenomenon in Case (CR). In this case, the Cold Bath extracts phonons out of the dot and $I_Q^{el}$ flows from the contacts to the dot. The whole scenario is pictorially presented in Fig.\ref{9e}. It is intriguing to notice  that $I_Q^{el}$ can be stimulated by varying the temperature of phonon reservoir even at the short circuit point $V=0$, where the effective electron flow vanishes (i.e $I=0$). Now if the contact $L$ is strongly coupled to the dot as compared to the contact $R$, (i.e $\gamma_L >> \gamma_R$), $I_Q^{el}$ is dragged out from (or pushed into) the dot by the contact $L$. Fig.\ref{9f} and \ref{9g} plot the small voltage response of $I_{QL}^{el}$ and $I_{QR}^{el}$ for non-zero $\lambda$ in Case (HR) and (CR) respectively. They show that the hotter (colder) reservoir can selectively heat up (or cool down) $L$ even when $V=0$ as shown in the schematic presented in Fig.\ref{9h}. Therefore, the temperature of phonon reservoir plays a major role in the stimulation and control of the electronic and phonon heat current in the dot. We believe that this intriguing physics can be implemented in the design of thermal transistor or thermal logic devices. \\
\section{\label{sec:level4} Conclusion} 
\indent The anomalous behavior of phonon transport due to finite electron-phonon interaction was investigated using an \textit{Anderson-Holstein} based dissipative quantum dot setup in two relevant cases: (a)  electron flow stimulated by a voltage bias in the absence of an electronic temperature gradient and (b) electron flow driven by the electronic temperature gradient at zero voltage. We explained the observed cumulative effects of voltage and electronic temperature gradients on the  non-linear phonon currents, using a new transport coefficient termed as \textit{electron induced phonon thermal conductivity}. It was demonstrated that under suitable operating conditions in Case (a) the dot  pumped in phonons into the hotter phonon reservoirs and in Case (b) the dot extracted phonons out of the colder phonon reservoirs. Finally, we elaborated how the non-linear electronic heat current was stimulated and controlled by engineering the temperature of the phonon reservoirs coupled to the dot.
	
\bibliographystyle{apsrev}

\bibliography{refrences}
\end{document}